\documentclass[twocolumn]{aastex63}

\usepackage{graphicx}
\usepackage[mathscr]{euscript}
\usepackage{xcolor}
\usepackage[normalem]{ulem}
\usepackage{hyperref}
\hypersetup{
 colorlinks,
citecolor = [rgb]{0,0.5,0} }
\usepackage{amsmath}
\usepackage{soul}
\usepackage{multirow}
\usepackage{lipsum}

\newcommand{\vect}[1]{\boldsymbol{#1}}

\def\gsim{\mathrel{\raise.3ex\hbox{$>$\kern-.75em\lower1ex\hbox{$\sim$}}}}
\def\lsim{\mathrel{\raise.3ex\hbox{$<$\kern-.75em\lower1ex\hbox{$\sim$}}}}
\def\leq{\mathrel{\raise.3ex\hbox{$<$\kern-.75em\lower1ex\hbox{$-$}}}}

\makeatletter
\newcommand\footnoteref[1]{\protected@xdef\@thefnmark{\ref{#1}}\@footnotemark}
\makeatother

\begin{document}

\title{Improved Methods for Estimating Peculiar Velocity Correlation Functions Using Volume Weighting}

\correspondingauthor{Yuyu Wang}
\email{yuyuwang@sjtu.edu.cn}

\author{Yuyu Wang}
\affiliation{Department of Astronomy, School of Physics and Astronomy, Shanghai Jiao Tong University, Shanghai, 200240, China.}
\affiliation{Department of Physics \& Astronomy, University of Kansas, Lawrence, KS 66045, USA.}

\author{Sarah Peery}
\affiliation{Department of Physics, Willamette University, Salem, OR 97301, USA.}

\author{Hume A. Feldman}
\affiliation{Department of Physics \& Astronomy, University of Kansas, Lawrence, KS 66045, USA.}

\author{Richard Watkins}
\affiliation{Department of Physics, Willamette University, Salem, OR 97301, USA.}


\begin{abstract}

We present an improved method for calculating the parallel and perpendicular velocity correlation functions directly from peculiar velocity surveys using weighted maximum-likelihood estimators. A central feature of the new method is the use of position-dependent weighting scheme that reduces the influence of nearby galaxies, which are typically overrepresented relative to the more distant galaxies in most surveys. We demonstrate that the correlation functions calculated this way are less susceptible to biases due to our particular location in the Universe, and thus are more easily comparable to linear theory and between surveys. Our results suggest that the parallel velocity correlation function is a promising cosmological probe, given that it provides a better approximation of a Gaussian distribution than other velocity correlation functions and that its bias is more easily minimized by weighting. Though the position weighted parallel velocity correlation function increases the statistical uncertainty, it decreases the cosmic variance and is expected to provide more stable and tighter cosmological parameter constraints than other correlation methods in conjunction with more precise velocity surveys in the future.

\end{abstract}


\section{INTRODUCTION}
\label{sec:introduction_1}

Studies of density perturbations provide information used to analyze the large scale structure of the Universe. However,  density perturbation studies based on redshift galaxy distributions are limited by the bias due to  peculiar velocities, also known as redshift space distortion (RSD). Many studies have shown the effects of peculiar velocities in RSD studies \citep[e.g.][]{Kaiser1987, MelColFelWil1998, ThoMelFelSha2004, Scoccimarro2004, TarNisSai2010, ReiWhi2011, SelMcD2011, ZhaPanZhe2013, ZheZhaJin2013, SonNisTar2013, TarnisBer2013, SenZal2014, UhlKop2015, OkuHanSel2015, VlaCasWhi2016, BiaPerBel2016, HanSelBeu2017, BelPezCar2018}.

Peculiar velocity is a powerful tracer of mass distribution \citep[e.g.][]{WatFelHud2009, FelWatHud2010, DavNusMas2011, NusBraDav2011, MacFelFer2011, Turnbull2012, MacFelFer2012, Nusser2014, SprMagCol2014, JohBlaKod2014, ScrDavBla2015}. However,  current peculiar velocity measurements are still based on  radial distances, which limit the precision of peculiar velocity surveys. A different method of measuring the peculiar velocity can be made using the kinematic Sunyaev-Zel'dovich effect \citep[e.g.][]{SunZel1980, DolHanRon2005, KasAtrKoc2008, HanAddAub2012, DolKomSun2015, PlanckXXXVII2015, PlanckAgh2019}. However, due to the signal weakness, it is a very difficult measurement. Therefore, ensemble statistics of peculiar velocities is more practical for current studies \citep[e.g.][]{Kaiser1988, FerJusFel1999, JusFerFelJaf2000, FelJusFer2003, WatFel2007, WatFelHud2009, FelWatHud2010, DavNusMas2011, AgaFelWat2012, AbaFel2012, HanAddAub2012, Nusser2014, Hellwing2014, PlanckXXXVII2015, KumWanFelWat2015, ScrDavBla2015, ScrDavBla2015, SeiPar2016, HofNusCorTul2016, Nusser2016, HelNusFei2017}. 

Velocity correlation function analysis provides another tool to investigate the peculiar velocity field. The most widely used velocity correlation estimator was introduced by \citet{Gorski1988} and further formulated in \citet{GorDavStr1989}. It has provided interesting results constraining cosmological parameters  \citep[e.g.][]{JafKai1995, ZarZehDekHof1997, JusFerFelJaf2000, BorCosZeh2000,  AbaErd2009, NusDav2011, OkuSelVla2014, HowStaBla2017, HelNusFei2017, WanRooFel2018, DupCouKub2019}. 

The velocity correlation function can be expressed as two independent functions, one for velocity components along the separation vector of a pair of galaxies and one for components perpendicular to this vector.   The \citet{Gorski1988} correlation estimator results in a complicated combination of these two functions, with the precise mixture given by selection functions that depend on the distribution of the survey objects as well as the separation distance.  This estimator has the decided disadvantage of not being comparable between studies that use different survey objects. Furthermore, at the time that it was introduced it was seen as being more stable than other methods given the small size of the available datasets.   Given the availability of much larger peculiar velocity catalogs today, it is an opportune time to explore other methods of estimating velocity correlations.   In addition, \citet{WanRooFel2018} found that the cosmic variance of the correlation function using the Gorski estimator is large and non-Gaussian distributed, and  \citep{HelNusFei2017} showed that it is susceptible to biases due to our special location near a large overdensity -- the Virgo Cluster.   These problems make the \citet{Gorski1988} peculiar velocity correlation estimator less than ideal as a probe of large-scale structure.   

In this paper, we use an alternative method, introduced by \cite{Kaiser1989} and \cite{GroJusOst1989}, that estimates the parallel and perpendicular correlation functions directly in a way that is independent of the survey distribution.   This method further allows for the weighting of individual velocity measurements to account for the uneven sampling of the volume by peculiar velocity surveys.  This is caused by two effects.  First, the density of observed galaxies in a survey typically decreases with distance, so that the inner portions of the survey volume are more densely sampled than the outer portions.  Second, peculiar velocity measurement uncertainties grow rapidly with distance, so that the measured velocities of nearby objects are much more accurate than those at greater distance.  Both of these effects result in nearby galaxies carrying an outsized weight in most velocity analyses, leading to results that predominantly reflect the velocity field in a much smaller effective volume than expected from the scale of the survey.  While \cite{GroJusOst1989} used weighting to reduce the effect of random errors, we introduce a novel weighting scheme which reduces cosmic variance and bias by increasing the effective volume probed by a survey.    

The paper is organized as follows: In section~\ref{sec:correlation}, we derive the weighted estimators for the parallel and perpendicular correlation functions. In section~\ref{sec:data}, we discuss the CosmicFlow-3 (CF3) catalog we analyze. In section~\ref{sec:simulation}, we introduce the N-body simulations and methods used for generating mock catalogs. In section~\ref{sec:result}, we show results for our method on both randomly centered  mock catalogs as well as those centered in environments similar to that of the Milky Way for several different weighting schemes. We also apply our methods to obtain estimates of the parallel and perpendicular correlation functions in the local Universe using data from the CF3 catalog. In section~\ref{sec:constraints}, we discuss the parameter constraining result using the weighted estimators. Section~\ref{sec:conclusion} concludes this paper.

\section{The Peculiar Velocity Correlation Estimator}
\label{sec:correlation}

The general form of the two-point velocity correlation tensor is 
\begin{equation}
\Psi_{ij}({\bf r}) = \langle v_i({\bf r_0})v_j({\bf r_0}+{\bf r})\rangle, 
\end{equation}
where $i$ and $j$ designate the cartesian components of the velocity and the average is over points separated by the vector ${\bf r}$.   Making the usual assumption that the velocity field is a statistically isotropic and homogeneous random field, we can write the correlation tensor in terms of two functions which depend only on the magnitude of the separation vector $r=|{\bf r}|$, 
\begin{equation}
\Psi_{ij}(r)= \Psi_\parallel(r){\hat r_i\hat r_j}+ \Psi_\perp(r)\left(\delta_{ij} - {\hat r_i\hat r_j}\right)
\label{eq:full_psi}
\end{equation}
where $\hat r$ is a unit vector in the direction of the separation vector.  These two functions have simple physical interpretations; $\Psi_\parallel(r)$ is the (parallel) correlation of the velocity components along the separation vector and $\Psi_\perp(r)$ gives the (perpendicular) correlation of the components of the velocity perpendicular to the separation vector.   

Our goal is to estimate $\Psi_\parallel(r)$ and $\Psi_\perp(r)$ from the correlations in the radial component of the peculiar velocity, $u$, which is the only component that can be measured.   
Given a pair of galaxies at positions ${\bf r_1}$ and ${\bf r_2}$, we can write the correlation of their radial peculiar velocities as

\begin{equation}
\begin{aligned}
\langle u_1u_2\rangle &=  \hat r_{1i}\hat r_{2j}\langle v_iv_j\rangle \\
&= \Psi_\parallel(r)({\bf \hat r_1\cdot \hat r)(\hat r_2\cdot \hat r}) \\
& + \Psi_\perp(r)\left[{\bf \hat r_1\cdot \hat r_2}-({\bf \hat r_1\cdot\hat r})({\bf \hat r_2\cdot \hat r})\right] .
\end{aligned}
\label{eq:scorr}
\end{equation}

This expression can be written in terms of $\theta_1$ and $\theta_2$, the angles the separation vector ${\bf r}$ makes with the position vectors ${\bf r_1}$ and ${\bf r_2}$ respectively.   Specifically, 
\begin{equation}
({\bf \hat r_1\cdot \hat r})({\bf \hat r_2\cdot \hat r})= \cos(\theta_1)\cos(\theta_2),
\end{equation}
and
\begin{equation}
{\bf \hat r_1\cdot \hat r_2}= \cos(\theta_2-\theta_1) = \cos(\theta_1)\cos(\theta_2) + \sin(\theta_1)\sin(\theta_2).
\end{equation}
Using these results, we can put Eq.~\ref{eq:scorr} into the simple form
\begin{equation}
\langle u_1u_2\rangle =  \Psi_\parallel(r)f(\theta_1,\theta_2)+ \Psi_\perp(r)g(\theta_1,\theta_2),
\end{equation}
where $f=\cos(\theta_1)\cos(\theta_2)$ and $g=\sin(\theta_1)\sin(\theta_2)$.

Following \cite{Kaiser1989} and \cite{GroJusOst1989}, we use a weighted least-squares method to estimate $\Psi_\parallel(r)$ and $\Psi_\perp(r)$ from a catalog of peculiar velocities $u_m$.  We minimize the function
\begin{equation}
\chi^2(r) = \sum_{m,n} w_{m,n}\left[ u_mu_n - \Psi_\parallel(r) f(\theta_1,\theta_2) - \Psi_\perp(r) g(\theta_1,\theta_2)\right]
\label{eq:chisq}
\end{equation}
with respect to $\Psi_\parallel(r)$ and $\Psi_\perp(r)$, where the sum is over pairs of galaxies whose separations fall within a specified bin and  $w_{i,j}$ is a weight assigned to each galaxy pair.     The minimization can be done analytically, resulting in the estimates
\begin{equation}
\Psi_{\parallel} (r)= \frac{\sum wg^2 \sum wf u_1u_2- \sum wfg \sum wg u_1u_2}{\sum wf^2 \sum wg^2 - \left( \sum wfg \right)^2 } ,
\label{eq:psipar}
\end{equation}
and
\begin{equation}
\Psi_{\perp}(r) = \frac{\sum wf^2 \sum wg u_1u_2 - \sum wfg \sum wf u_1u_2}{\sum wf^2 \sum w g^2 - \left( \sum wfg \right)^2 },
\label{eq:psiperp}
\end{equation} 
where the sums are over galaxy pairs whose separations lie in a bin centered on $r$.   

An alternative approach to studying peculiar velocity correlations is to use the $\psi_1$ and $\psi_2$ statistics introduced by \cite{GorDavStr1989} and utilized in several subsequent studies \cite[e.g.][]{BorCosZeh2000, HelNusFei2017, WanRooFel2018}.  While these statistics in principle carry the same information as $\Psi_\parallel$ and $\Psi_\perp$, in practice they depend on the particular distribution of objects in a survey, making them not comparable between surveys.   While in the past there was some motivation to focus on $\psi_1$ as being particularly stable when applied to the small datasets available at the time, there is now sufficient data to estimate $\Psi_\parallel$ and $\Psi_\perp$ directly.   It is possible to calculate $\Psi_\parallel$ and $\Psi_\perp$ from $\psi_1$ and $\psi_2$ given the positions of the the survey objects \cite[see e.g.][]{WanRooFel2018}; however, this process can be shown to be mathematically equivalent to the calculations shown in Eqs.~\ref{eq:psipar} and ~\ref{eq:psiperp}.

It is not obvious how to best  choose weights to use in Eqs.~\ref{eq:chisq}, \ref{eq:psipar} and ~\ref{eq:psiperp}.  \cite{Kaiser1989} used the simplest choice,  $w=1$, while \cite{GroJusOst1989} chose weights with an eye towards reducing the effects of measurement errors.  However, previous work \citep{WanRooFel2018} has shown that, for the surveys we are working with, statistical errors are small compared to the effects of cosmic variance, since we estimate the correlation function in a volume that is smaller than the scale of homogeneity. In general, the number density of galaxies nearby is larger than the number density of galaxies in the distant volume, this problem is exacerbated by the large measurement uncertainty of distant galaxies. This ``concentration" of galaxies at small distances puts greater emphasis on the nearby volume, so that the effective volume reflected in the correlation functions can be significantly smaller than that of the survey.   This effect increases the cosmic variance and may also lead to bias.   Here we will weight the pairs in order to better ``balance" the survey, so that it has a larger effective volume and hence smaller cosmic variance and bias but may lead to larger statistical errors.   

Our approach will be to weight pairs of galaxies by the factor $w=(r_1r_2)^p$, where $r_1$ and $r_2$ are the distances to the galaxies and $p$ is a positive power.     This scheme gives less weight to pairs of nearby galaxies, which are overrepresented in the sample, and greater weight to pairs of more distant galaxies, which are underrepresented.   Correlation functions calculated using this weighting should thus sample the volume of the survey more evenly, and hence reflect a larger effective volume.  However, in giving greater weight to galaxies that are far away, and hence have larger peculiar velocity uncertainties, our weighting scheme will necessarily increase statistical errors.  We will explore several different choices for the power $p$ in order to determine which value provides the best overall statistic for the data we are working with.   

When analyzing data from simulations, we have access to all three components of the peculiar velocity.  In this case we can calculate $\Psi_\parallel$ and $\Psi_\perp$ directly by taking a weighted average of products of velocity components parallel and perpendicular to the separation vector for each pair, namely
\begin{equation}
\label{eq:3d_1}
\Psi^{3D}_{\parallel}(r) = \sum_{pairs}w\left( \vect{v_1}\cdot\vect{r} \right) \left( \vect{v_2}\cdot\vect{r} \right)/ \sum_{pairs}w 
\end{equation}
and
\begin{equation}
\label{eq:3d_2}
\Psi^{3D}_{\perp} = \frac{1}{2} \sum_{pairs} w\left[ (\vect{v_1}\cdot \vect{v_2}) - \left( \vect{v_1}\cdot\vect{r} \right) \left( \vect{v_2}\cdot\vect{r} \right) \right]/\sum_{pairs}w 
\end{equation}
where ${\bf r}= {\bf r_2}-{\bf r_1}$.

In linear theory, $\Psi_\parallel$ and $\Psi_\perp$ can be related directly to the power spectrum of density fluctuations $P(k)$ \citep{EisHu1998} through the relations 
\begin{eqnarray}
\label{eq:psi_para}
\Psi_\parallel(r)  &=& \frac{\left( \sigma_8 f H_0\right) ^2}{2\pi^2\sigma^2(8)} \int P(k)\left[j_0(kr) - 2\frac{j_1(kr)}{kr}\right]dk\, ,\\
\label{eq:psi_perp}
\Psi_\bot(r) &=& \frac{\left( \sigma_8 f H_0\right) ^2}{2\pi^2\sigma^2(8)} \int P(k)\frac{j_1(kr)}{kr} dk ,
\end{eqnarray}
where $f=\Omega^{0.55}_m$ \citep{Linder2005}, $H_0$ is the Hubble constant, $j_n(x)$ are the spherical Bessel functions, $\sigma_8$ is the amplitude of density fluctuations on a scale of 8 $h^{-1}$Mpc. In the equations,  $\sigma_8$ is the value from the simulation we use (see Section~\ref{sec:simulation}) and the $\sigma(8)$ is calculated following the method in \citet{EisHu1998} (Eq. A7).

\section{Data}
\label{sec:data}

The CosmicFlows-3 (CF3) peculiar velocity compilation \citep{CF3} includes two catalogs: the galaxy catalog and the group catalog. The CF3-galaxy catalog contains 17,669 galaxies, including all the 8,135 CosmicFlows-2 (CF2) \citep{TulCouDol2013} galaxy distances, which is  a compilation of Type Ia Supernovae (SNIa) \citep{TonSchBar}, Spiral Galaxy Clusters (SC) TF clusters \citep[]{GioHaySal1998, DalGioHay1999}, Streaming Motions of Abell Clusters (SMAC) FP clusters \citep{HudSmiLuc1999, HudSmiLuc2004}, Early-type Far Galaxies (EFAR) FP clusters \citep{ColSagBur2001}, TF clusters \citep{Willick1999}, the SFI++ catalog \citep{MasSprHay2006, SprMasHay2007, SprMasHay2009}, group SFI++ catalog \citep{SprMasHay2009}, Early-type Nearby Galaxies (ENEAR) survey \citep{daCBerAlo2000, BerAlodaC2002, WegBerWil2003}, and a surface brightness fluctuations (SBF) survey \citep{TonDreBla2001}, together with 2,257 distances derived from the correlation between galaxy rotation and luminosity with photometry at 3.6$\mu m$ obtained with Spitzer Space Telescope and 8,885 distances based on the Fundamental Plane sample derived from the Six Degree Field Galaxy Survey (6dFGS) \citep{SprMagCol2014}. The CF3-group catalog contains 11,878 groups and galaxies, where galaxies in known groups have had their distance moduli and redshifts averaged, resulting in a single velocity and position for the group as a whole.  Due to this averaging, peculiar velocities of groups have reduced uncertainties compared to individual galaxies.  However, \citet{DupCouKub2019} suggests that using grouped data to constrain the growth rate might lead to incoherent results. In the following analyses, we will use  the CF3-galaxy catalog.  

The peculiar velocities of the CF3 are calculated through the unbiased peculiar velocity estimator introduced by \citet{WatFel2015a}:
\begin{equation}
\label{peculiar_v}
v = cz\log\left( \frac{cz}{H_0 r}\right) .
\end{equation}
The redshift ($cz$) and distance ($r$) are provided by the CF3 survey, however, the choice of the value of Hubble constant will affect the peculiar velocity and therefore also the velocity correlation result. \citet{WanRooFel2018} discussed the effect of the Hubble constant on the 
\citet{Gorski1988} correlation functions. For this study we will set the Hubble constant equal to 75 km s$^{-1}$ Mpc$^{-1}$ for the peculiar velocities of CF3 survey;  \citep{CF3} have shown that this is the value that minimizes the magnitude of radial flows.   

Due to  large uncertainties in distance measurements, previous studies of the velocity correlation functions \citep[e.g.][]{Gorski1988, BorCosZeh2000, WanRooFel2018} used redshifts to determine positions of objects and hence the separations between them. 
Distances given by $cz/H_o$ differ from the actual distances by an ``error" of the peculiar velocity divided by the Hubble constant, which can be much smaller than the measurement uncertainty for measured distances.  
In \citet{WanRooFel2018} we found that using the estimated distances to estimate the galaxy pair separation leads to unreliable $\psi_1$ and $\psi_2$ results. Considering the relations among $\psi_1$, $\psi_2$, $\Psi_{\parallel}$ and $\Psi_{\perp}$, redshift is also the optimal choice for the $\Psi_{\parallel}$ and $\Psi_{\perp}$ separations. Therefore, even though using redshift as the separation may lead to redshift distortion effects, it is still more reliable than biases caused by the large uncertainty of distance estimation. In this paper we will also use redshifts to determine positions of the objects in our catalog, using distance estimates only in our calculation of peculiar velocities.

\section{Mock catalogs}
\label{sec:simulation}

The mock catalogs we use in this paper are generated from halo catalogs of the OuterRim Simulation \citep{HabPopFin2016, HeiFinPop2019, HeiUraFin2019}, which is carried out from the Mira-Titan Universe Simulations. The OuterRim simulation is a dark matter only simulation with cosmological parameters similar to the WMAP-7 \citep{WMAP7} cosmology, which are shown in Table~\ref{T_Mill}.

\begin{table}[!ht]
\caption{The cosmological parameters of the OuterRim Simulation}
\centering
{
\begin{tabular}{lc}
\hline
Matter density, $\Omega_m$ & 0.2648\\
Cosmological constant density, $\Omega_\Lambda$ & 0.7352 \\
Baryon density, $\Omega_b$ & 0.0448\\
Hubble parameter, $h$ (100 km s$^{-1}$ Mpc$^{-1}$) & 0.71\\
Amplitude of matter density fluctuations, $\sigma_8$ & 0.8\\
Primordial scalar spectral index, $n_s$ & 0.963\\
Box size ($h^{-1}$Gpc) & 3.0\\
Number of particles & $10,240^3$\\
Particle mass, $m_p$ ($10^{9} h^{-1} M_{\odot}$) & 1.85\\
Softening, $f_c$ ($h^{-1}$kpc) & 3\\
\hline
\end{tabular}%
}
\label{T_Mill}
\end{table}

\begin{figure}[!ht]
\centering
\includegraphics[width=8.5cm]{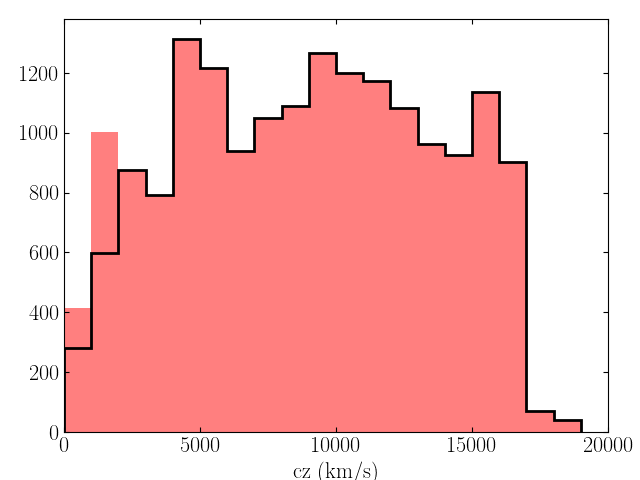}
\caption{The redshift distributions of the full CF3-galaxy catalog(red histogram). The black lined histogram shows the example of the mock catalogs. }
\label{fig:radialdist}
\end{figure}

The simulation contains halos with a large range of masses that cover galaxies, groups, and clusters. We use halos in the mass range [$10^{11}$, $10^{13}$]$M_\odot$ located in the inner (1.5 $h^{-1}$Gpc)$^3$ volume of the OuterRim Simulation box as galaxies to generate mock catalogs that mimic the CF3-galaxy survey. Figure \ref{fig:radialdist} shows the redshift distribution of the CF3-galaxy catalog and an example of one mock catalog. We found that choosing mock catalog galaxies to match the CF3 selection function based on redshifts or distances had no significant impact on the correlation function.   As discussed above, we chose the redshift selection function in generating mock catalogs since, given the large uncertainty in distance estimates,  redshift provides a more accurate distance.   

We generated two kinds of mock catalogs for the error analysis of the correlation function: catalogs for cosmic variance and catalogs for statistical errors. The cosmic variance mocks ($M_c^i$) centered at randomly distributed positions without any peculiar velocity measurement uncertainties, where $M_c^i$ represents the cosmic variance mock locates at the $i^{th}$ center. The cosmic variance is calculated by taking standard deviations of the velocity correlation functions of the cosmic variance mocks ($\sigma_c = STD\left\lbrace M_c^1, M_c^2, \cdots, M_c^{100}\right\rbrace $). We used two types of cosmic variance mock catalogs differing in how their halo center points are chosen.  The first type, which we call ``random" mocks, are centered on randomly chosen halos inside the the inner (1.5 $h^{-1}$Gpc)$^3$ region of the simulation.  The second type, called Local Group (LG)  mocks, are centered on a Milky Way-like halo ($ M=\left[13.5 \pm 6.5 \right] \times 10^{11} h^{-1} M_{\odot}$]) with a Virgo-like cluster at a similar distance of the Virgo Cluster from the Milky Way.   The LG selection criterion is based on that introduced by \citet{HelNusFei2017}.  These catalogs are useful for exploring the observational consequences of our non-typical location in the Universe, a region whose dominant characteristic is the neighboring Virgo-like cluster ($M=\left[1.2 \pm 0.6 \right] \times 10^{15} h^{-1} M_{\odot}$) at a distance of $12\pm4$ $h^{-1}$Mpc. 

For each type of  observers, we generated 100 mock catalogs. With mock catalog centers selected randomly, it would be hard to avoid overlapping between them. However, the overlapping is not significant in such a large volume. For random observers, only 7.6\% of the pair distance of mock centers are closer than 300 $h^{-1}$Mpc to each other. For LG observers, 7\% of them are closer than 200 $h^{-1}$Mpc and 15\% of them are less than 300 $h^{-1}$Mpc. Considering the maximum depth of CF3-galaxy survey is about 170 $h^{-1}$Mpc (Figure~\ref{fig:radialdist}) and the galaxy separations of interest for this study are smaller than 100 $h^{-1}$Mpc, the  overlapping is negligible. 

\begin{figure}[!ht]
\centering
\includegraphics[width=8.5cm]{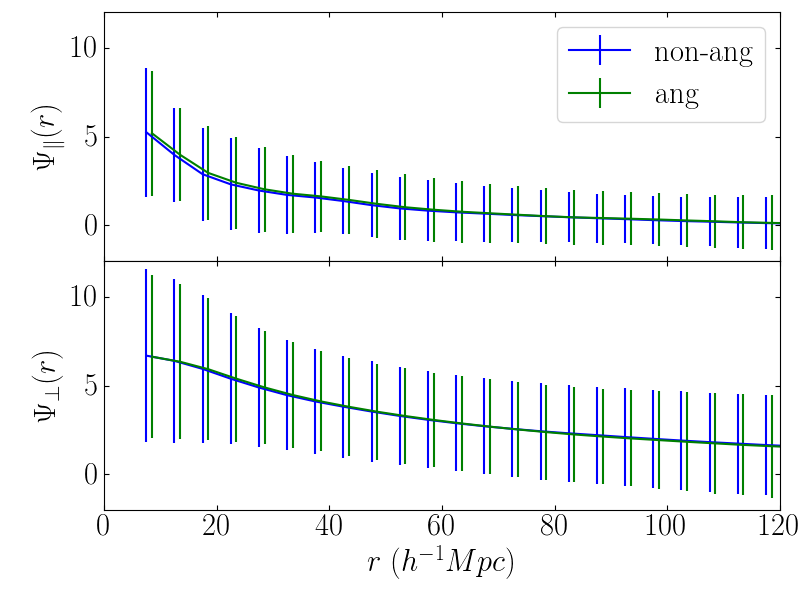}
\caption{$\Psi_{\parallel}$ and $\Psi_{\perp}$ results of mock catalogs with (green) and without (blue) the anisotropy of CF3 angular distribution. The error bars show the total error of the correlations.}
\label{fig:ang_compare}
\end{figure}

We also generated mock catalogs that mimic the angular distribution of the CF3 objects, which is significantly anisotropic.  However, we found that the anisotropy of the CF3 angular distribution does not have a significant effect on the correlation functions, as shown in Figure~\ref{fig:ang_compare}.  

The statistical error mock catalogs ($M_{s_j}^i$) are generated by perturbing the distances (and hence peculiar velocities) of the objects in a cosmic variance mock catalog with the average CF3 distance measurement uncertainty (20\%), where $M_{s_j}^i$ represents the $j^{th}$ perturbed statistical error mock catalog at the $i^{th}$ center. The statistical error of the $i^{th}$ cosmic variance mock catalog is calculated by taking standard deviations over 100 versions of the statistical error mock catalogs ($\sigma_s^i = STD\left\lbrace M_{s_1}^i, M_{s_2}^i, \cdots, M_{s_{100}}^i \right\rbrace $), while the statistical error of the velocity correlation function is calculated by taking average of the statistical errors over 10 randomly selected cosmic variance mock catalogs ($\sigma_s = AVE\left\lbrace \sigma_s^i \right\rbrace_{10}, i\in[1,100] $).

\section{Results}
\label{sec:result}

\begin{figure}[!ht]
\centering
\includegraphics[width=8.5cm]{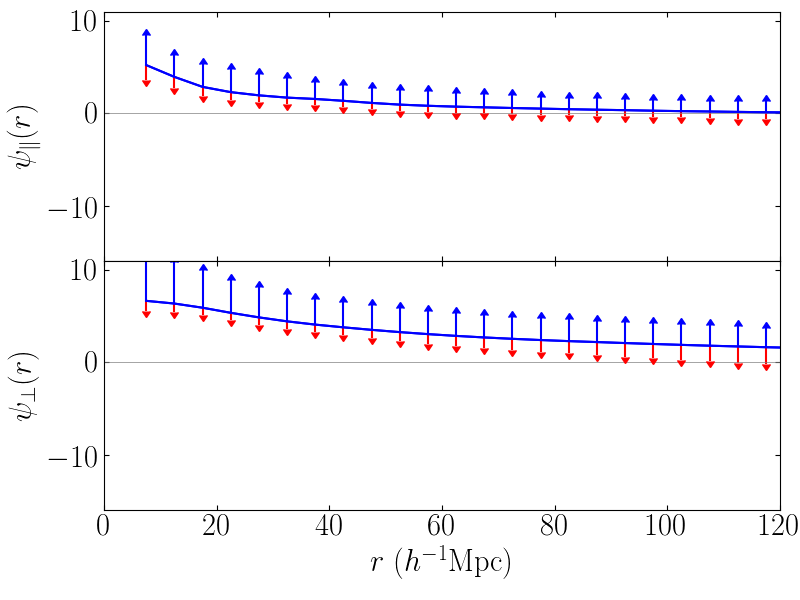}
\caption{\label{fig:cv_st} The parallel and perpendicular correlation functions of randomly centered mock catalogs with uniform weighting. $\Psi_{\parallel}$ and $\Psi_{\perp}$ are in units of (100 km s$^{-1}$)$^2$. The blue solid lines show the average values for 100 mock catalogs. The upper blue error bars show the cosmic variance. The lower red error bars indicate the statistical error.}
\end{figure}
 
\begin{figure}[!ht]
\centering
\includegraphics[width=8.5cm]{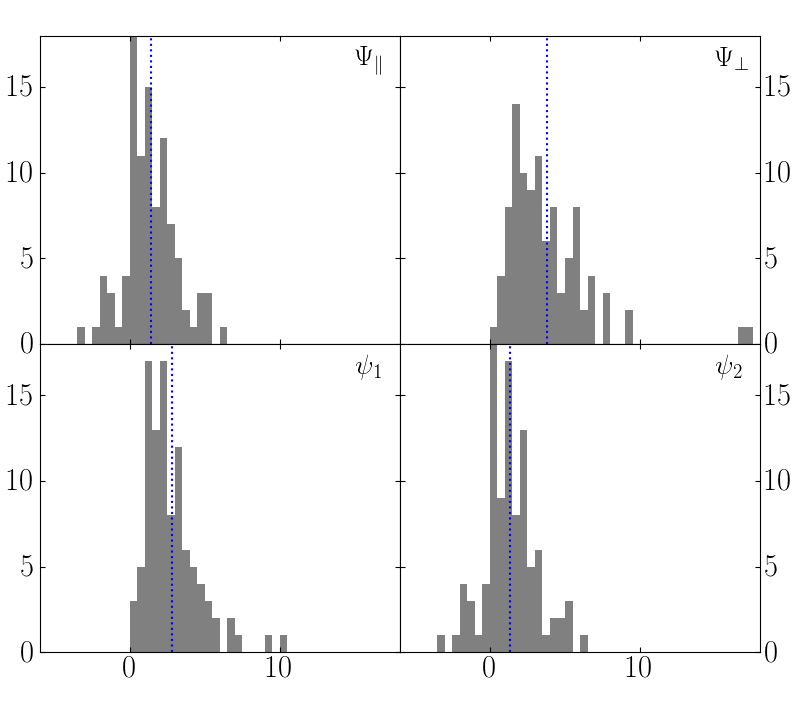}
\caption{\label{fig:cv-fitting} The distribution of $\Psi_{\parallel}$, $\Psi_{\perp}$, $\psi_1$, and $\psi_2$ in 40-45 $h^{-1}$Mpc bin of 100 randomly centered mock catalogs using uniform weighting, in units of (100 km s$^{-1}$)$^2$. The blue dotted vertical line is the mean of the mock catalogs.}
\end{figure}

Figure~\ref{fig:cv_st} shows $\Psi_{\parallel}$ and $\Psi_{\perp}$ and their cosmic variance (upper) and statistical errors (lower) using randomly centered mock catalogs with uniform weighting. In the figure, the cosmic variance, which is larger than the statistical errors (especially for closer pairs), dominates the error budget. This is consistent with the \citet{WanRooFel2018} results that showed that the cosmic variance is the dominant source of error in the Gorski correlation functions, which use uniform weighting.  \citet{WanRooFel2018} also showed that  the error distribution of the function $\psi_1$ was significantly non-Gaussian. Below we will examine the question of the distribution of the correlation functions in more detail.   

\begin{figure}[!ht]
\centering
\includegraphics[width=8.5cm]{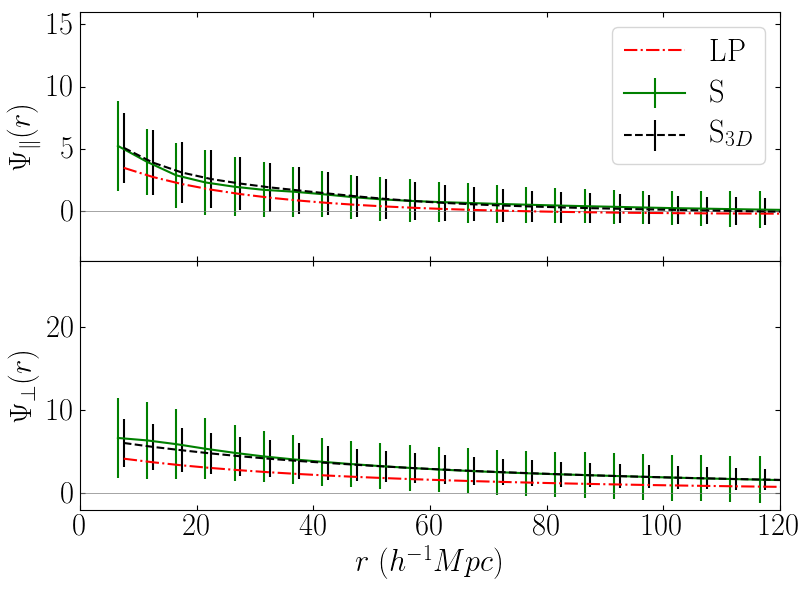}
\caption{\label{fig:rand_noang} The parallel and perpendicular correlation functions of 100 randomly centered mock catalogs in units of (100 km s$^{-1}$)$^2$. The red dash-dotted lines show the linear predictions (LP). The green solid lines indicate the average of mock catalog results calculated using the estimators discussed in the text (S). The black dashed lines indicate the average of mock catalog results for the full 3D velocity fields ($S_{3D}$). The error bars show the cosmic variance of the mock catalogs.}
\end{figure}

Figure~\ref{fig:cv-fitting} shows the cosmic variance distribution of $\Psi_{\parallel}$ and $\Psi_{\perp}$ calculated from our estimators using uniform weighting, and $\psi_1$, and $\psi_2$ calculated using the \citet{Gorski1988} formalism, for 100 randomly centered mock catalogs.  We show the distributions for a particular bin (40-45 $h^{-1}$Mpc) as an example. In the figure, we see that  $\Psi_{\parallel}$ and $\psi_2$ have roughly Gaussian distributions, while the distributions of $\Psi_{\perp}$ and $\psi_1$ are noticeably skewed, with significant non-Gaussian tails;  these distributions generally become more non-Gaussian in larger separation bins. The similarity of $\Psi_{\parallel}$ and $\psi_2$ is not surprising, since $\psi_2$ is calculated from the projections of the radial velocities onto the separation vectors.  The other Gorski correlation function, $\psi_1$, is estimated from the unprojected radial velocity, making it a combination of $\Psi_{\parallel}$ and $\Psi_{\perp}$.  
The cosmic variance of $\Psi_{\parallel}$ is roughly Gaussian except for scales smaller than 10 $h^{-1}$Mpc, where the uncertainty of the correlation function is large. Considering the large uncertainty and possible non-Gaussian cosmic variance in small separation bins, we recommend that small scale correlations ($\lsim$ 10 $h^{-1}$Mpc) not be used in parameter constraints. Quantities with non-Gaussian distributions are difficult to interpret, suggesting that studies of the velocity correlation function should focus on $\Psi_{\parallel}$. We will return to this issue in Section~\ref{sec:conclusion}. 

Figure~\ref{fig:rand_noang} shows the $\Psi_{\parallel}$ and $\Psi_{\perp}$ estimators (Eqs.~\ref{eq:psipar} and~\ref{eq:psiperp}) and 3D velocity fields (Eqs.~\ref{eq:3d_1} and~\ref{eq:3d_2})  calculated using randomly centered mock catalogs. The simulation results agree well with linear predictions for both $\Psi_{\parallel}$ and $\Psi_{\perp}$. Although the estimators use only line of sight peculiar velocities, they also agree well with the full 3D results, lending credence to their efficacy and stability. It should be noted again that there is the potential of redshift distortion effects between the mock catalog results and the linear theory prediction, since the correlation function is calculated with redshift separations while the linear prediction is calculated from distance separations. However, out results indicate that these effects are not significant.   

\begin{figure}[!ht]
\centering
\includegraphics[width=8.5cm]{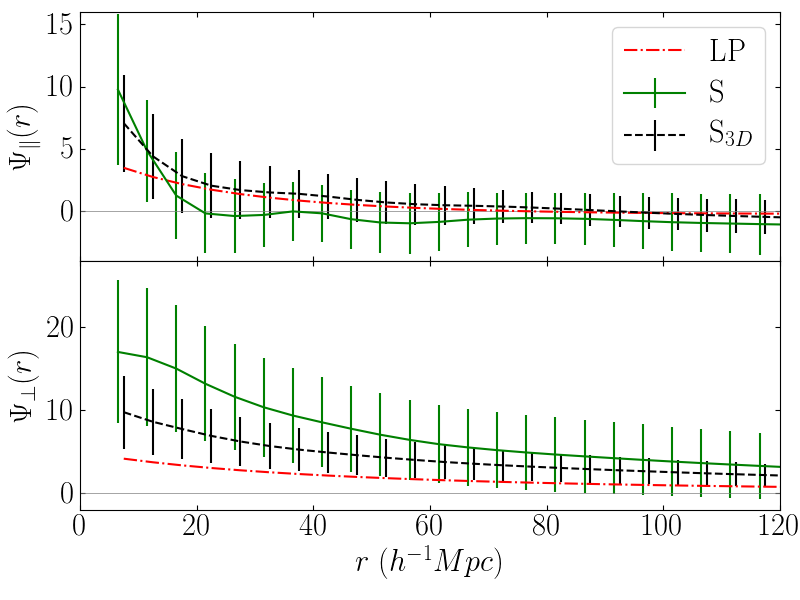}
\caption{\label{fig:LG_noang} Same as Figure~\ref{fig:rand_noang} but using LG centered mock catalogs.}
\end{figure}

\citet{HelNusFei2017} discussed the effect of observer location on velocity statistics.  They compared both the Gorski velocity correlation function estimators and the pairwise velocity statistic calculated for mock catalogs with random halo centers and for those centered on locations that mimicked the local group (LG).   They found that the correlation functions calculated from the local group-like catalogs exhibited significant bias relative to linear theory.  To study the effects of the observer location on the parallel and perpendicular correlation functions, in Figure~\ref{fig:LG_noang} we show the results of using LG centered mock catalogs.  We see that $\Psi_{\parallel}$ and $\Psi_{\perp}$ for the LG centered mocks with uniform weighting are also biased.  As we discuss  below, this bias can be greatly reduced through the use of weighting.  

Figure~\ref{fig:LG_noang} shows the parallel and perpendicular correlation results of using LG centered mock catalogs with uniform weighting ($w=1$).   We see that the restriction to local group-like locations introduces significant systematic bias into our results relative to linear theory.   This bias takes two distinct forms.  First, we see that both our estimators, which use only radial velocities, do not accurately recover the 3D correlation function.   Second, we see that, especially for the perpendicular correlation function, the average correlations calculated from the 3D velocities also do not accurately reflect linear theory.   Both of these biases arise, most likely, because the volumes around the LG centered mocks are not ``typical", but rather exhibit particular flow patterns that are significantly different than the averages taken over random volumes.     

\begin{figure}[!ht]
\centering
\includegraphics[width=8.5cm]{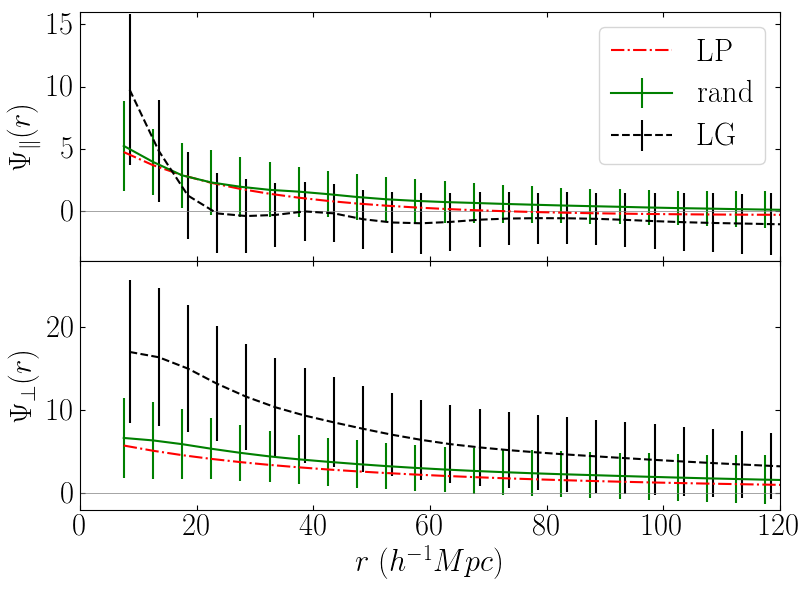}
\caption{\label{fig:psi_rand_LG} Parallel and perpendicular correlation results of 100 random centered and 100 LG centered mocks in units of (100 km s$^{-1}$)$^2$ using uniform weighting ($w=1$). All mock catalogs have had galaxy distances perturbed by random measurement errors.   The red dash-dotted lines show the linear predictions. The green solid lines indicate the average results for the randomly centered mocks. The black dashed lines show the average results for the LG centered mocks. The error bars show the total error of the correlation function, which includes both cosmic variance and statistical error.}
\end{figure}

Figure~\ref{fig:psi_rand_LG} shows a comparison between random and LG centered mocks. The error bars of the simulation results show the total error ($\sigma_t = \sqrt{\sigma_c^2+\sigma_s^2}$, where $\sigma_t$ is the total error, $\sigma_c$ is the cosmic variance and $\sigma_s$ is the statistical error) of the correlation functions.  We see that the variance of the LG centered mock catalogs is significantly larger than that of the randomly centered mock catalogs, particularly for the perpendicular correlation function ($\Psi_{\perp}$).  

The fact that the bias in the estimated correlation functions with uniform weights in LG centered mocks has the same order of magnitude as the correlation functions themselves suggests that correlation functions calculated using the CF3 with uniform weights, which includes those calculated using the Gorski method, should not be used in comparisons with linear theory.   

As discussed above, weighting can be used to increase the effective volume of the survey.   Our approach will be to weight galaxy pairs by $w=(r_1r_2)^p$, where $r_1$ and $r_2$ are the distances of the two galaxies and $p$ is a non-negative power.   In the LG centered mocks, this weighting reduces emphasis on the relatively small volume near the center of the survey, which for the LG centered mocks is atypical.   We will see that the use of weighting can effectively reduce the bias found in the LG centered mocks.   

\begin{figure*}
\centering
\includegraphics[width=5.5cm]{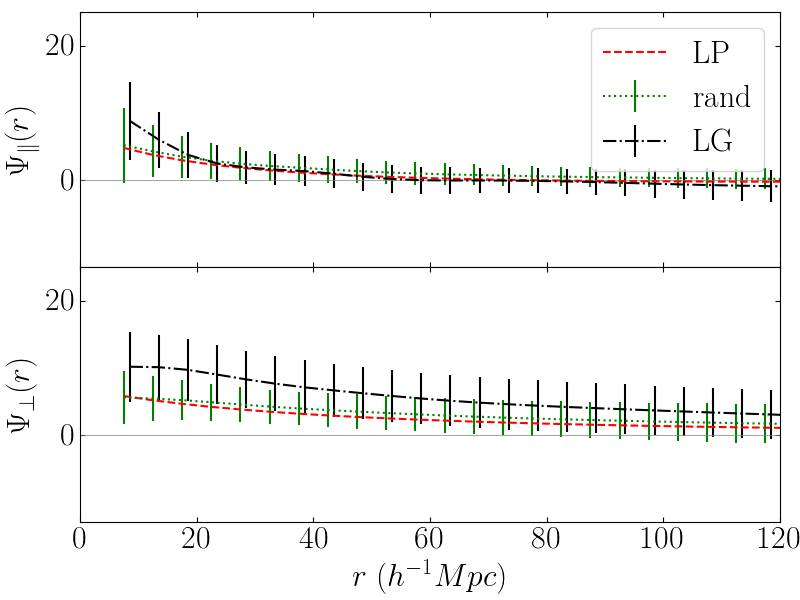}
\includegraphics[width=5.5cm]{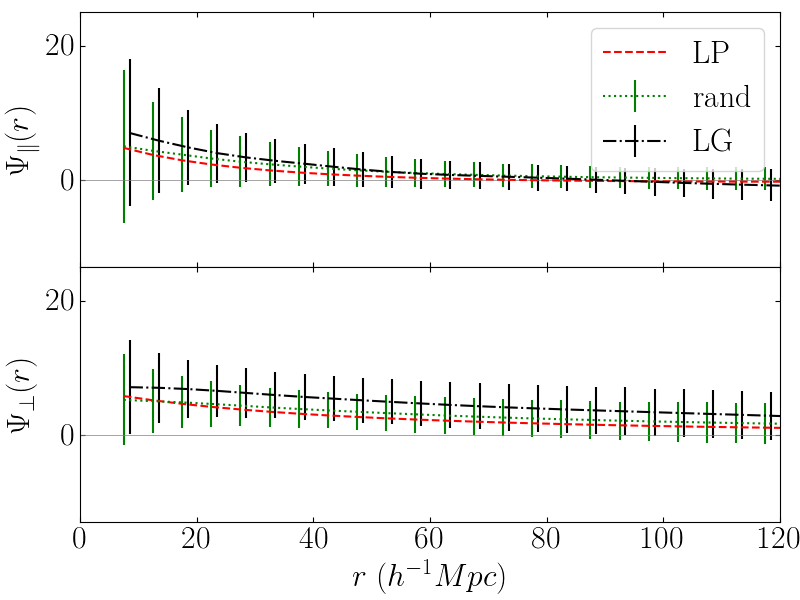}
\includegraphics[width=5.5cm]{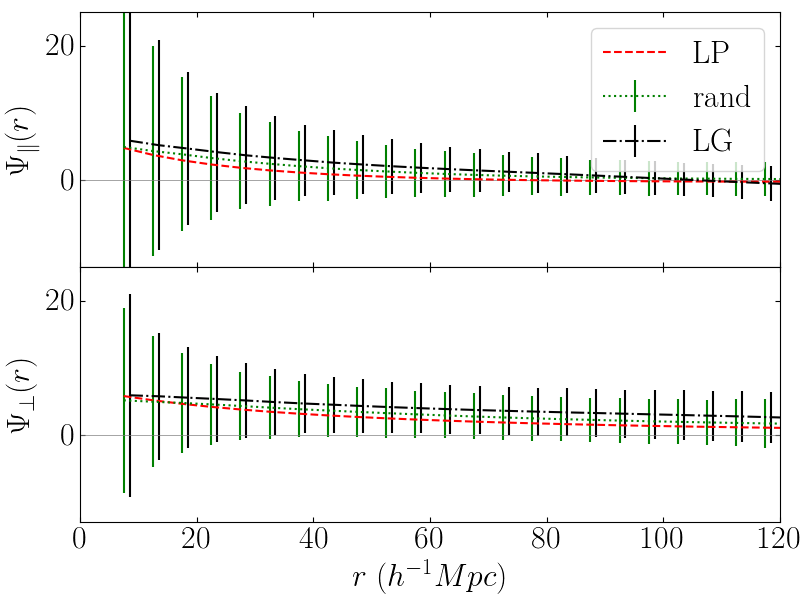}
\caption{\label{fig:weights} Same as Figure~\ref{fig:psi_rand_LG} but using position weighted method with weights $p=1/2$ (left panel), $p=1$ (middle panel), and $p=2$ (right panel).}
\end{figure*}

Figure~\ref{fig:weights} shows the results using weights $w=(r_1r_2)^p$ with $p=0.5,1,2$, respectively ($p=0$ gives uniform weights).   The use of weighting has reduced the bias to an insignificant level.    However, the total error becomes larger while increasing the effective volume of the surveys. 

\begin{figure*}[!ht]
\centering
\includegraphics[width=5.5cm]{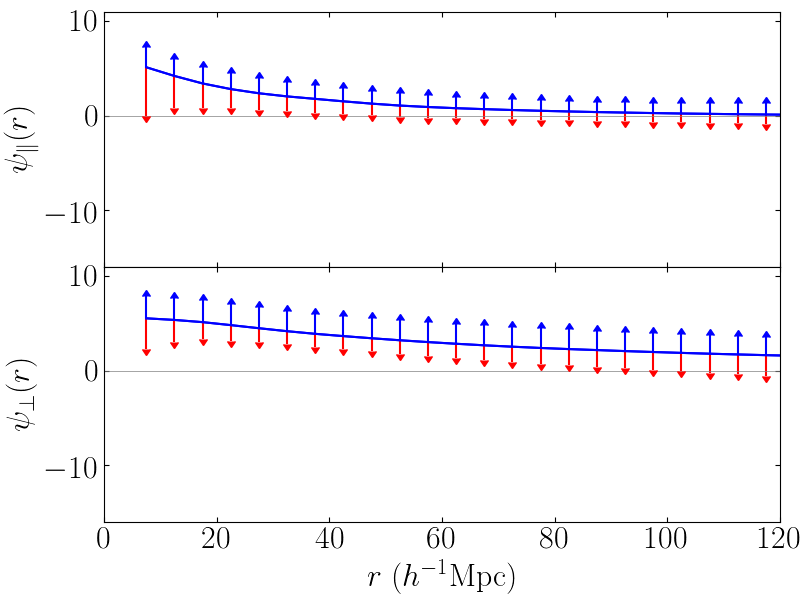}
\includegraphics[width=5.5cm]{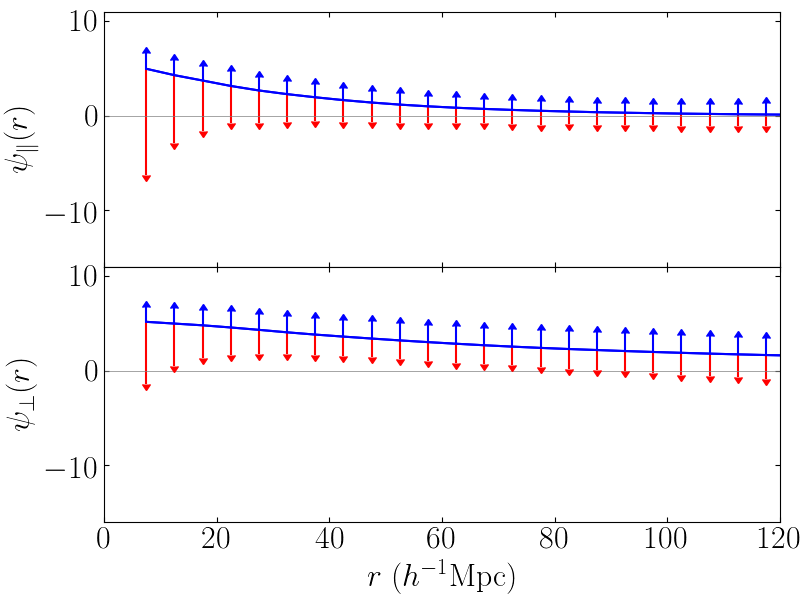}
\includegraphics[width=5.5cm]{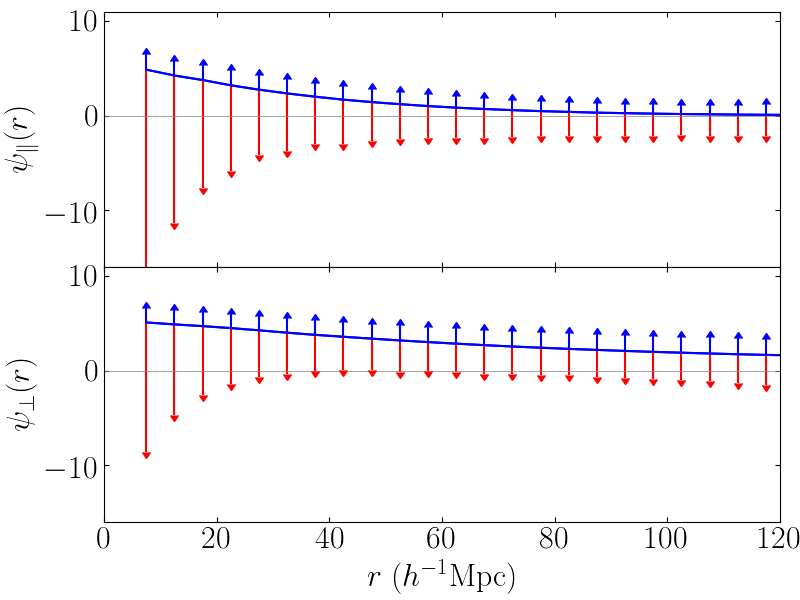}
\caption{\label{fig:cv_st_weights} The parallel and perpendicular correlation functions with weights $p=\frac{1}{2}$ (left panel), $p=1$ (middle panel), and $p=2$ (right panel). $\Psi_{\parallel}$ and $\Psi_{\perp}$ are in units of (100 km s$^{-1}$)$^2$. The blue solid lines show averages over 100 mock catalogs. The upper blue error bars show the cosmic variance. The lower red error bars indicate the statistical errors.}
\end{figure*}

Figure~\ref{fig:cv_st_weights} shows the cosmic variance and the statistical errors of the weighted correlation functions with $p=\frac{1}{2}$ ($w=(r_1r_2)^{1/2}$), $p=1$ ($w=r_1r_2$), and $p=2$ ($w=(r_1r_2)^2$), respectively. The cosmic variance generally decreases with weighting as expected; however, the statistical errors increase and dominate when using weights with larger $p$. In addition, both our result and result from \citet{HelNusFei2017} show that the cosmic variance of LG observers is larger than that of random observers. \citet{HelNusFei2017} explanation is that different observers see different radial velocity components for the same galaxies. 

Considering the decreasing trend of the cosmic variance with the weight power, we  suggest that the large cosmic variance of LG observers may also be caused by the imbalanced (nearby galaxies dominated) distribution of galaxies . When the galaxy distribution is imbalanced, the galaxy distribution around a big attractor (e.g. Virgo Cluster) may lead to various biases and large cosmic variance. Table~\ref{tb:weight_t} shows the cosmic variance and statistical errors of $\Psi_{\parallel}$ with different weighting schemes; even as cosmic variance decreases, we see the total error increase. Considering the tradeoffs, $p=1$ ($w=r_1r_2$) seems to be the best value to use for the CF3 survey. However, the optimal value of $p$ may vary for different surveys due to the specific object distributions and uncertainties.

\begin{table}[!ht]
\caption{The errors of $\Psi_{\parallel}$ with random observer using various weighting schemes. The $\sigma_t$, $\sigma_c$ and $\sigma_s$ indicate the total, cosmic variance and statistical errors of the correlation function in units of (100 km s$^{-1}$)$^2$, respectively. Numbers in the brackets show the separation range of three selected bins (in units of $h^{-1}$Mpc).}
\centering
{
\begin{tabular}{cccrrr}
\hline
 Weight &$p$& & [15-20] & [35-40] & [55-60]\\
 \hline
& & $\sigma_c$ & 2.45 & 1.88 & 1.59 \\
1 &0& $\sigma_s$ & 0.95 & 0.66 & 0.66 \\
 & & $\sigma_t$ & 2.63 & 1.99  & 1.72 \\
 \hline
 & & $\sigma_c$ & 1.78 & 1.48 & 1.29 \\
 $\sqrt{r_1r_2}$& $\frac12$ & $\sigma_s$ & 2.58 & 1.45 & 1.13 \\
 & & $\sigma_t$ & 3.13 & 2.07 & 1.71 \\
 \hline
 & & $\sigma_c$ & 1.59 & 1.37 & 1.14 \\
 $r_1r_2$& 1 & $\sigma_s$ & 5.37 & 2.54 & 1.79 \\
 &  & $\sigma_t$ & 5.6 & 2.89 & 2.12 \\
 \hline
 & & $\sigma_c$ & 1.59 & 1.47 & 1.27 \\
 $(r_1r_2)^2$&2& $\sigma_s$ & 11.39 & 5.04 & 3.36 \\
 & & $\sigma_t$ & 11.5 & 5.25 & 3.59 \\
 \hline
\end{tabular}
}
\label{tb:weight_t}
\end{table}

\begin{figure}[!ht]
\centering
\includegraphics[width=8.5cm]{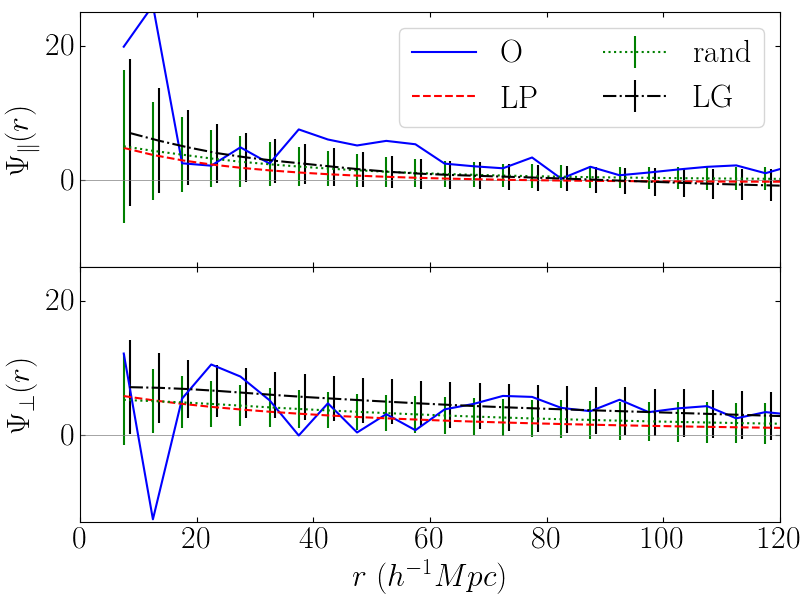}
\caption{\label{fig:observation_r1r2_s} The blue dotted lines indicate the parallel and perpendicular correlation estimates in units of (100 km s$^{-1}$)$^2$ calculated from the CF3-galaxy catalog using weighting scheme $w=r_1r_2$ ($p=1$).  The red dash-dotted lines show the linear prediction. The green solid lines indicate the average results from randomly centered mock catalogs with the same weighting. The black dashed lines show the average results from the LG centered mocks, also with the same weighting. The error bars show the total uncertainty, including cosmic variance and measurement errors.}
\end{figure}

Now that we have determined that $p=1$ provides the optimal weighting for our analysis, we apply our methods to the actual CF3-galaxy catalog. In Figure~\ref{fig:observation_r1r2_s}, we show the parallel and perpendicular correlation functions for the CF3-galaxy catalog, using the $p=1$ ($w=r_1r_2$) weighting scheme, together with the results (with estimated total uncertainties, including cosmic variance and measurement errors) of both the random and LG centered mock catalogs with the same weighting.   We see that both $\Psi_\parallel$ and $\Psi_{\perp}$ have the expected behavior: decreasing amplitude with increasing separation.  Also as expected from linear theory, $\Psi_{\perp}$ decreases more slowly and has larger amplitude than $\Psi_\parallel$ at large separation.   Considering the magnitudes of the total uncertainties, both $\Psi_\parallel$ and $\Psi_{\perp}$ are consistent (within two standard deviations) with the results from the mock catalogs, and thus consistent with the standard cosmological model.

\section{Parameter Constraints}
\label{sec:constraints}

In \citet{WanRooFel2018}, we showed that the correlation function $\psi_1$ has a long, non-Gaussian tail in its cosmic variance distribution, making it unsuitable for placing constraints on cosmological parameters.  As we discussed in section~\ref{sec:result}, the cosmic variance of $\psi_{\parallel}$ exhibits a better approximation of a Gaussian distribution than $\psi_1$.  This suggests that $\psi_{\parallel}$ may be a more useful measurement of peculiar velocity correlations.  
In this section, we test the performance of uniformly weighted and position weighted $\psi_{\parallel}$ with both random and LG observers with respect to putting constraints on cosmological parameters.

As can be seen in Figure~\ref{fig:cv_st_weights}, the statistical errors increase with weighting even as the cosmic variance decreases. To gauge the effects of statistical errors and cosmic variance separately and together, we implement three methods for cosmological parameter estimation.  To look at the effects of cosmic variance alone, we use mock catalogs drawn from different regions of the simulation box with no measurement errors in their velocities (as described above) with the $\chi^2$ given by 
\begin{equation}
\chi_c^2 = \sum_{i,j}\left[ \psi^C_{\parallel}(r_i) - \psi^L_{\parallel}(r_i)\right] C^{-1}_{ij}\left[ \psi^C_{\parallel}(r_j) - \psi^L_{\parallel}(r_j)\right],
\label{eq:convariance_m}
 \end{equation}
where
 \begin{equation}
C_{ij} = \frac{1}{N_{mock}} \sum^{N_{mock}}_{l=1} \left( \Psi^i_{\parallel ,l} - \overline{\Psi}^i_{\parallel, C} \right) \left( \Psi ^j_{\parallel ,l} - \overline{\Psi}^j_{\parallel, C}\right).
\end{equation}

$\chi^2_c$ (Eq.~\ref{eq:convariance_m}) is the cosmic variance ($C$) covariance matrix, $N_{mock}=100$ is the number of mock catalogues; $ \psi^i_{\parallel, l}$ is the correlation value of the $i^{th}$ separation bin of the $l^{th}$ mock catalogue; $\overline{\psi}^i_{\parallel, C}$ is the average value of $N_{mock}$ catalogues in the $i$th separation bin; $\psi^C_{\parallel}$ is the average value of $\psi_{\parallel}$ over $N_{mock}$ mock catalogs; $\psi^L_{\parallel}$ is the linear prediction.

\begin{figure*}[!ht]
\centering
\includegraphics[width=17cm]{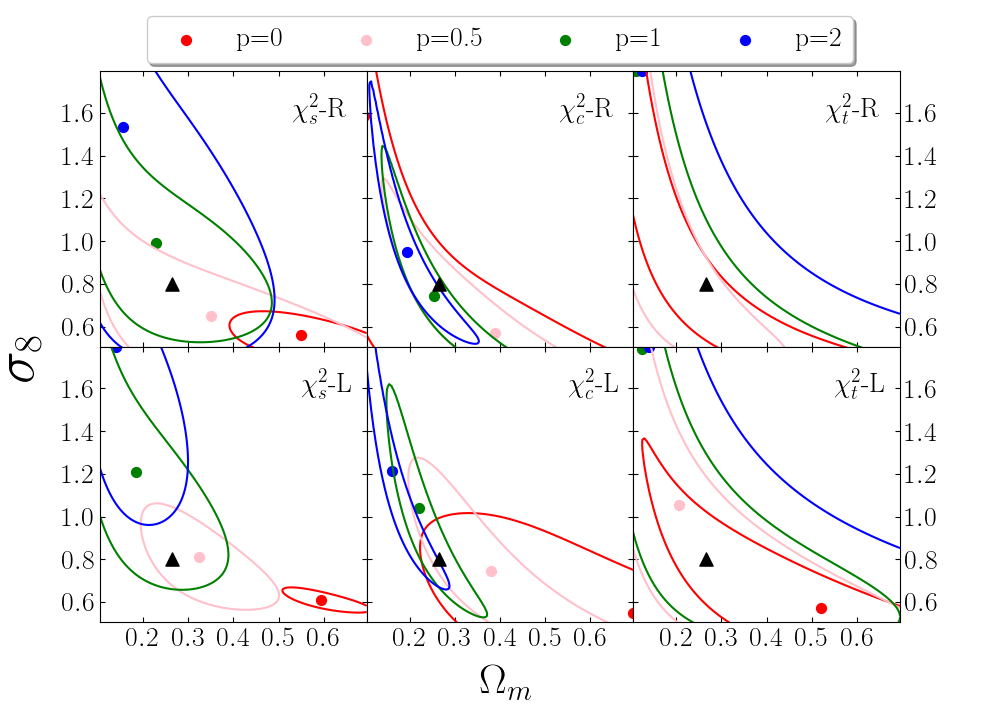}
\caption{\label{fig:parameter}  $\Omega_m$ and $\sigma_8$ constrains using simulation data with bin equals $500$ km s$^{-1}$ within separation scales $[1000, 6000]$ km s$^{-1}$. The minimum $\chi^2$ value has been subtracted from each cell. The contours indicate 68\%  likelihood of $\chi^2$ values. The triangle marker indicates the value from the OuterRim Simulation. $\chi^2_c$ shows the result of covariance matrix with cosmic variance, $\chi^2_s$ indicates the result of covariance matrix with statistical errors, and $\chi^2_t$ includes the information of both the cosmic variance and statistical errors. R and L indicate the random and LG observers, respectively.}
\end{figure*} 

To look at the effects of measurement errors alone, we use 100 versions of one mock catalog perturbed with artificial measurement errors with the $\chi^2$ function
 \begin{equation}
 \resizebox{0.5\textwidth}{!}{$\chi_s^2 = \sum_{i,j}\left[ \psi^S_{\parallel}(r_i) - \psi^L_{\parallel}(r_i)\right]\varepsilon^{-1}_{ij} \left[ \psi^S_{\parallel}(r_j) - \psi^L_{\parallel}(r_j)\right],$}
\label{eq:convariance_e}
\end{equation}
where
 \begin{equation}
 \varepsilon_{ij} = \frac{1}{N_{pert}} \sum^{N_{pert}}_{p=1} \left( \Psi^i_{\parallel, p} - \Psi^i_{\parallel, A} \right) \left( \Psi ^j_{\parallel, p} - \Psi^j_{\parallel, A}\right).
\end{equation}

$\chi^2_s$ (Eq.~\ref{eq:convariance_e}) uses the covariance matrix that contains the information of statistical errors, where $\varepsilon$ is the covariance matrix of statistical errors; $N_{pert}$ is the number of perturbed catalogs of a selected mock catalog whose value is closest to the average value of the 100 mock catalogs; $\psi^i_{\parallel, A}$ is the parallel correlation of the selected mock catalog in the $i$th separation bin; $ \psi^i_{\parallel, p}$ is the correlation value of the $i^{th}$ separation bin of the $p^{th}$ perturbed catalog of the selected mock catalog; $\psi^S_{\parallel}$ is the average value over the $N_{pert}$ perturbed catalogs of the selected mock catalog.

Finally, for looking at the effects of cosmic variance and statistical errors together, we use 100 mock catalogs drawn from different parts of the simulation box, each perturbed with measurement errors.  The $\chi^2$ for these catalogs is given by 
 \begin{equation}
 \resizebox{0.5\textwidth}{!}{$\chi_t^2 = \sum_{i,j}\left[ \psi^T_{\parallel}(r_i) - \psi^L_{\parallel}(r_i)\right] T_{ij}^{-1}  \left[ \psi^T_{\parallel}(r_j) - \psi^L_{\parallel}(r_j)\right],$}
\label{eq:convariance_t}
\end{equation}
where
\begin{equation}
T_{ij} = \frac{1}{N_{mp}} \sum^{N_{mp}}_{n=1} \left( \Psi^i_{\parallel, n} - \overline{\Psi}^i_{\parallel, T} \right) \left( \Psi ^j_{\parallel, n} - \overline{\Psi}^j_{\parallel, T}\right).
\end{equation}

$\chi^2_t$ (Eq.~\ref{eq:convariance_e}) includes the effect of both the cosmic variance and statistical errors, where $T$ is the covariance matrix of total error; $N_{mp}$ is the number of perturbed mock catalogues, which means perturbing each of the 100 mock catalogs one time randomly according to distance uncertainties to get 100 perturbed mock catalogs; $ \psi^i_{\parallel, n}$ is the correlation value of the $i^{th}$ separation bin of the $n^{th}$ perturbed mock catalog; $\overline{\psi}^i_{\parallel, T}$ is the average value of $N_{mp}$ perturbed mock catalogs in the $i$th separation bin; $\psi^T_{\parallel}$ is the average value over the $N_{mp}$ perturbed mock catalogs.

\begin{figure*}[!ht]
\centering
\includegraphics[width=17cm]{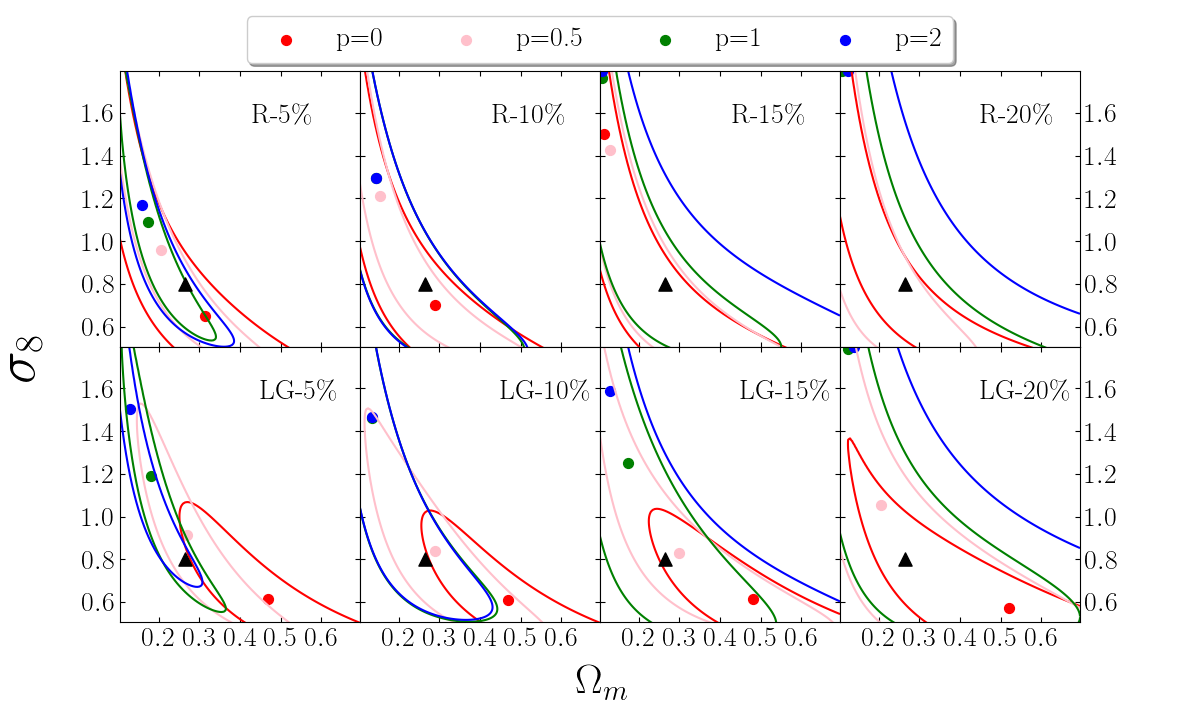}
\caption{\label{fig:Chi_t}  $\Omega_m$ and $\sigma_8$ constrains of $\chi^2_t$ with different distance perturbation values. The minimum $\chi^2_t$ value has been subtracted from each cell. The contours indicate 68\%  likelihood values. The triangle marker indicates the value from the OuterRim Simulation. R and LG indicate the random and local group observers, respectively.}
\end{figure*} 

Figure~\ref{fig:parameter} shows the cosmological parameters constraints for $\Omega_m$ and $\sigma_8$ using $\Psi_{\parallel}$ in separation scales $[1000, 6000]$ km s$^{-1}$ with bin width equal $500$ km s$^{-1}$, which are used consistently in the following parameter constraints. In our tests, we find the results using $\Psi_{\parallel}$ are much more stable than using $\psi_1$ for implementing different truncations \citep[see also][]{WanRooFel2018}.

For the $\chi^2_c$ fitting method, all of the four correlation weights, $p=0$ ($w=1$), $p=0.5$ ($w=(r_1r_2)^{1/2}$), $p=1$ ($w=r_1r_2$), $p=2$ ($w=(r_1r_2)^2$), agree with the simulation value within $1\sigma$ for both the random and LG observers. However, the results of the uniform weighted $\Psi_{\parallel}$ with the LG observer are not as consistent as the results of random observers. The position weighted method improves the parameter constraints for the LG observer significantly, since the position weighting scheme reduces the bias. The position weighting scheme also provides tighter and more stable constraints than the uniform weighted $\Psi_{\parallel}$. In addition, the position weighted $\Psi_{\parallel}$ provides tighter constraints to the expected value (simulation value) for both the random and LG observers. Comparing the results of the three position weighted $\Psi_{\parallel}$, $p=1$ provides better results.

In the $\chi^2_s$ plots, the result of uniform weighted $\Psi_{\parallel}$ biases from the simulation value for both random and LG observers. However, the position weighted correlation function agrees with the simulation value within $1\sigma$ for both types of observers, except $p=2$ for the LG observer. Similar to the results of the $\chi^2_c$ method, the uniform weighted $\Psi_{\parallel}$ provides a biased parameter constraints for the LG observer, which is greatly improved by the position weighting scheme. However, unlike $\chi^2_c$, the position weighted $\Psi_{\parallel}$ has laxer constraining contours than the uniform weighted $\Psi_{\parallel}$. This is due to the larger statistical errors caused by the larger position weighting power. 
 

To study the effect of the size of the statistical uncertainty on the $\chi^2_t$ constraining method, we implement different distance uncertainty percentages (distance uncertainty equal to $5\%$, $10\%$, $15\%$, $20\%$ of distance) as shown in Figure~\ref{fig:Chi_t}. In the figure, the position weighted correlations show significant improvements on LG observers for all of the four different uncertainty percentages. However, the $\chi^2_t$ constraining contour becomes large when the distance perturbation is larger than $10\%$. Therefore a CF3-like survey with 20\% distance errors will probably not be able to put meaningful constraints on cosmological parameters.  

Much larger peculiar velocity surveys should be available in the not-too-distant future. 
Having more survey objects will improve constraints in two main ways.   First, since the correlation function is essentially an average, having more survey objects will reduce statistical errors in the usual way.  However, having more survey objects, particularly at large distances, will also allow us to reduce cosmic variance by using a more aggressive weighting scheme and therefore increasing the effective volume that the survey probes.  In other words, if the statistical errors start smaller, then we can afford to have them increase more in our effort to reduce cosmic variance.  Without weighting, increasing the number of survey objects without substantially increasing survey depth will have less effect, since statistical errors are currently dominated by cosmic variance.  
As the size of peculiar velocity surveys increases, our method should allow us to use peculiar velocity correlations to place significant constraints on cosmological parameters. 

\begin{figure}[!ht]
\centering
\includegraphics[width=8.5cm]{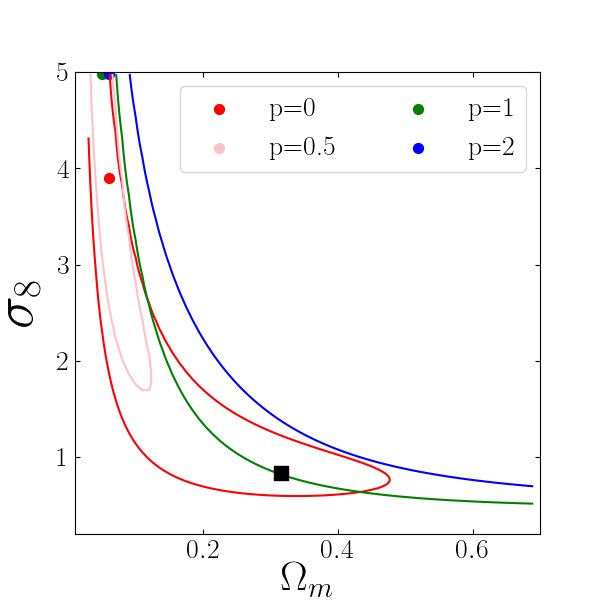}
\caption{\label{fig:Chi_observ}  $\Omega_m$ and $\sigma_8$ constraints obtained from the CF3-galaxy survey using $\chi^2_t$ with the covariance matrix calculated using mock LG catalogs. 
The contours indicate 68\%  likelihood values. The square marker indicates the best value from Planck \citep{Planckparameters2014} .}
\end{figure}

The $\chi^2_t$ is the appropriate one to use for real data, it accounts for both cosmic variance and measurement errors.  
In Figure~\ref{fig:Chi_observ} we show the constraining result obtained by applying our method to the CF3-galaxy survey.  In our calculations, we use the covariance matrix calculated from mock catalogs of LG observers with 20\% distance uncertainties; this covariance matrix should be the best match to the cosmic variance and measurement uncertainties of the real data.  
In the figure, the observation constraining results agree with Planck \citep{Planckparameters2014} within 1$\sigma$, except for the $p=0.5$. However, as expected the constraining contour is wide and flat and thus does not constrain the values significantly.  In order to put tighter constraints on cosmological parameters we  require larger and/or more precise peculiar velocity surveys.  

We also tested $f\sigma_8$ constraints for the CF3-galaxy survey and found no improvements. Furthermore, the $f\sigma_8$ statistic is an approximation that may lead to loss of information. In Figure~\ref{fig:psi_rand_LG}, the bias caused by LG observers is mainly represented in the shape of the correlation function rather than its magnitude. The shape of the linear predicted correlation function is determined by the integral of the power spectrum (Eq.~\ref{eq:psi_para}), which requires the value of $\Omega_m$. $f\sigma_8$ constraining requires a fixed $\Omega_m$ value for the power spectrum, which leads to a fixed shape of the correlation function. This would defeat our purpose of reducing the bias caused by LG observers, since the bias is mainly represented by the shape of the correlation function.

\section{Conclusion}
\label{sec:conclusion}

Previous studies of velocity correlations have mostly focused on $\psi_1$, a correlation function introduced by Gorski \citep{Gorski1988}.  This function has several disadvantages. First, it is dependent on the distribution of objects being analyzed and hence is not comparable between surveys.  Second, it is a complicated mixture of the physically meaningful correlation functions that quantify correlations of the velocity components parallel and perpendicular to the separation vector between pairs of galaxies.   Third, as shown by \citet{WanRooFel2018}, the distribution of cosmic variance in $\psi_1$ is significantly nonGaussian, complicating its use as a cosmological probe.  
Finally, as noted by \citet{HelNusFei2017}, and as we have shown here, our special location near the Virgo cluster can bias correlation functions calculated using typical catalogs whose density of objects decreases rapidly with distance.

In this paper we have presented an alternative method, an extension of a method introduced by  \cite{Kaiser1989} and \cite{GroJusOst1989}, that can stably estimate the parallel and perpendicular correlation functions directly from currently available peculiar velocity data.   We have shown that the nonGaussian distribution of the cosmic variance in $\psi_1$ is mostly due to its containing $\Psi_{\perp}$; the parallel correlation function $\Psi_\parallel$ has a more Gaussian distribution and therefore should be much more useful as a cosmological statistic.  

We showed that the parallel and perpendicular correlation functions calculated with uniform weights are biased in LG centered mock catalogs, especially for small separations. The LG  mock catalog results also showed less agreement between the results of using estimators ($\Psi_\parallel$ and $\Psi_\perp$) and the results of using the full 3D velocity fields ($\Psi_\parallel^{3D}$ and $\Psi_\perp^{3D}$). $\Psi_{\perp}$ shows more bias, which explains the different behaviors shown by $\psi_1$ and $\psi_2$ when LG centered mock catalogs were used in \citet{HelNusFei2017}.  

Our results, together with those of  \citet{HelNusFei2017}, suggest that velocity correlation functions calculated from peculiar velocity data dominated by nearby galaxies will be biased due to our location near the Virgo Cluster.   We have presented a novel way to reduce this bias by including position weights into our analysis.  These weights reduce the emphasis on nearby galaxies, which are overrepresented in most catalogs. The weighted correlation functions probe a larger effective volume and thus give better agreement with linear theory.  In particular, we have shown that the bias due to our locations near the Virgo cluster is reduced when weights are used.  However, we find that there is a tradeoff between decreasing cosmic variance and increasing measurement uncertainties.  The optimal power to use will depend on the particular characteristics of the survey being analyzed.  

We find that the position weighted $\Psi_\parallel$ is a better cosmological probe than the previously used $\psi_1$ in that it has more Gaussian distributed errors.   While currently available peculiar velocity data is sufficient for calculating $\Psi_\parallel$ in the local Universe, it does not allow us to put significant constraints on cosmological parameters.  However, with larger and/or more accurate peculiar velocity surveys on the horizon, we expect velocity correlations to become an important cosmological probe.

\section{Acknowledgements}

HAF and RW were partially supported by NSF grant AST-1907404. An award of computer time was provided by the INCITE program. RW and SP acknowledge support from the Murdock Charitable Trust College Research Program.  

\bibliography{Yuyu}{}

\begin{thebibliography}{}
\expandafter\ifx\csname natexlab\endcsname\relax\def\natexlab#1{#1}\fi
\providecommand{\url}[1]{\href{#1}{#1}}
\providecommand{\dodoi}[1]{doi:~\href{http://doi.org/#1}{\nolinkurl{#1}}}
\providecommand{\doeprint}[1]{\href{http://ascl.net/#1}{\nolinkurl{http://ascl.net/#1}}}
\providecommand{\doarXiv}[1]{\href{https://arxiv.org/abs/#1}{\nolinkurl{https://arxiv.org/abs/#1}}}

\bibitem[{{Abate} \& {Erdo{\v g}du}(2009)}]{AbaErd2009}
{Abate}, A., \& {Erdo{\v g}du}, P. 2009, \mnras, 400, 1541,
  \dodoi{10.1111/j.1365-2966.2009.15561.x}

\bibitem[{{Abate} \& {Feldman}(2012)}]{AbaFel2012}
{Abate}, A., \& {Feldman}, H.~A. 2012, \mnras, 419, 3482,
  \dodoi{10.1111/j.1365-2966.2011.19988.x}

\bibitem[{{Agarwal} {et~al.}(2012){Agarwal}, {Feldman}, \&
  {Watkins}}]{AgaFelWat2012}
{Agarwal}, S., {Feldman}, H.~A., \& {Watkins}, R. 2012, \mnras, 424, 2667,
  \dodoi{10.1111/j.1365-2966.2012.21345.x}

\bibitem[{{Bel} {et~al.}(2019){Bel}, {Pezzotta}, {Carbone}, {Sefusatti}, \&
  {Guzzo}}]{BelPezCar2018}
{Bel}, J., {Pezzotta}, A., {Carbone}, C., {Sefusatti}, E., \& {Guzzo}, L. 2019,
  \aap, 622, A109, \dodoi{10.1051/0004-6361/201834513}

\bibitem[{{Bernardi} {et~al.}(2002){Bernardi}, {Alonso}, {da Costa}, {Willmer},
  {Wegner}, {Pellegrini}, {Rit{\'e}}, \& {Maia}}]{BerAlodaC2002}
{Bernardi}, M., {Alonso}, M.~V., {da Costa}, L.~N., {et~al.} 2002, \aj, 123,
  2990, \dodoi{10.1086/340463}

\bibitem[{{Bianchi} {et~al.}(2016){Bianchi}, {Percival}, \&
  {Bel}}]{BiaPerBel2016}
{Bianchi}, D., {Percival}, W.~J., \& {Bel}, J. 2016, \mnras, 463, 3783,
  \dodoi{10.1093/mnras/stw2243}

\bibitem[{{Borgani} {et~al.}(2000){Borgani}, {da Costa}, {Zehavi},
  {Giovanelli}, {Haynes}, {Freudling}, {Wegner}, \& {Salzer}}]{BorCosZeh2000}
{Borgani}, S., {da Costa}, L.~N., {Zehavi}, I., {et~al.} 2000, \aj, 119, 102,
  \dodoi{10.1086/301154}

\bibitem[{{Colless} {et~al.}(2001){Colless}, {Saglia}, {Burstein}, {Davies},
  {McMahan}, \& {Wegner}}]{ColSagBur2001}
{Colless}, M., {Saglia}, R.~P., {Burstein}, D., {et~al.} 2001, \mnras, 321,
  277, \dodoi{10.1046/j.1365-8711.2001.04044.x}

\bibitem[{{da Costa} {et~al.}(2000){da Costa}, {Bernardi}, {Alonso}, {Wegner},
  {Willmer}, {Pellegrini}, {Rit{\'e}}, \& {Maia}}]{daCBerAlo2000}
{da Costa}, L.~N., {Bernardi}, M., {Alonso}, M.~V., {et~al.} 2000, \aj, 120,
  95, \dodoi{10.1086/301449}

\bibitem[{{Dale} {et~al.}(1999){Dale}, {Giovanelli}, {Haynes}, {Campusano}, \&
  {Hardy}}]{DalGioHay1999}
{Dale}, D.~A., {Giovanelli}, R., {Haynes}, M.~P., {Campusano}, L.~E., \&
  {Hardy}, E. 1999, \aj, 118, 1489, \dodoi{10.1086/301048}

\bibitem[{{Davis} {et~al.}(2011){Davis}, {Nusser}, {Masters}, {Springob},
  {Huchra}, \& {Lemson}}]{DavNusMas2011}
{Davis}, M., {Nusser}, A., {Masters}, K.~L., {et~al.} 2011, \mnras, 413, 2906,
  \dodoi{10.1111/j.1365-2966.2011.18362.x}

\bibitem[{{Dolag} {et~al.}(2005){Dolag}, {Hansen}, {Roncarelli}, \&
  {Moscardini}}]{DolHanRon2005}
{Dolag}, K., {Hansen}, F.~K., {Roncarelli}, M., \& {Moscardini}, L. 2005,
  \mnras, 363, 29, \dodoi{10.1111/j.1365-2966.2005.09452.x}

\bibitem[{{Dolag} {et~al.}(2016){Dolag}, {Komatsu}, \&
  {Sunyaev}}]{DolKomSun2015}
{Dolag}, K., {Komatsu}, E., \& {Sunyaev}, R. 2016, \mnras, 463, 1797,
  \dodoi{10.1093/mnras/stw2035}

\bibitem[{{Dupuy} {et~al.}(2019){Dupuy}, {Courtois}, \&
  {Kubik}}]{DupCouKub2019}
{Dupuy}, A., {Courtois}, H.~M., \& {Kubik}, B. 2019, \mnras, 486, 440,
  \dodoi{10.1093/mnras/stz901}

\bibitem[{{Eisenstein} \& {Hu}(1998)}]{EisHu1998}
{Eisenstein}, D.~J., \& {Hu}, W. 1998, \apj, 496, 605, \dodoi{10.1086/305424}

\bibitem[{{Feldman} {et~al.}(2003){Feldman}, {Juszkiewicz}, {Ferreira},
  {Davis}, {Gazta{\~n}aga}, {Fry}, {Jaffe}, {Chambers}, {da Costa}, {Bernardi},
  {Giovanelli}, {Haynes}, \& {Wegner}}]{FelJusFer2003}
{Feldman}, H., {Juszkiewicz}, R., {Ferreira}, P., {et~al.} 2003, \apjl, 596,
  L131, \dodoi{10.1086/379221}

\bibitem[{{Feldman} {et~al.}(2010){Feldman}, {Watkins}, \&
  {Hudson}}]{FelWatHud2010}
{Feldman}, H.~A., {Watkins}, R., \& {Hudson}, M.~J. 2010, \mnras, 407, 2328,
  \dodoi{10.1111/j.1365-2966.2010.17052.x}

\bibitem[{{Ferreira} {et~al.}(1999){Ferreira}, {Juszkiewicz}, {Feldman},
  {Davis}, \& {Jaffe}}]{FerJusFel1999}
{Ferreira}, P.~G., {Juszkiewicz}, R., {Feldman}, H.~A., {Davis}, M., \&
  {Jaffe}, A.~H. 1999, \apjl, 515, L1, \dodoi{10.1086/311959}

\bibitem[{{Giovanelli} {et~al.}(1998){Giovanelli}, {Haynes}, {Salzer},
  {Wegner}, {da Costa}, \& {Freudling}}]{GioHaySal1998}
{Giovanelli}, R., {Haynes}, M.~P., {Salzer}, J.~J., {et~al.} 1998, \aj, 116,
  2632, \dodoi{10.1086/300652}

\bibitem[{{Gorski}(1988)}]{Gorski1988}
{Gorski}, K. 1988, \apjl, 332, L7, \dodoi{10.1086/185255}

\bibitem[{{Gorski} {et~al.}(1989){Gorski}, {Davis}, {Strauss}, {White}, \&
  {Yahil}}]{GorDavStr1989}
{Gorski}, K.~M., {Davis}, M., {Strauss}, M.~A., {White}, S.~D.~M., \& {Yahil},
  A. 1989, \apj, 344, 1, \dodoi{10.1086/167771}

\bibitem[{{Groth} {et~al.}(1989){Groth}, {Juszkiewicz}, \&
  {Ostriker}}]{GroJusOst1989}
{Groth}, E.~J., {Juszkiewicz}, R., \& {Ostriker}, J.~P. 1989, \apj, 346, 558,
  \dodoi{10.1086/168038}

\bibitem[{{Habib} {et~al.}(2016){Habib}, {Pope}, {Finkel}, {Frontiere},
  {Heitmann}, {Daniel}, {Fasel}, {Morozov}, {Zagaris}, \&
  {Peterka}}]{HabPopFin2016}
{Habib}, S., {Pope}, A., {Finkel}, H., {et~al.} 2016, \na, 42, 49,
  \dodoi{10.1016/j.newast.2015.06.003}

\bibitem[{{Hand} {et~al.}(2017){Hand}, {Seljak}, {Beutler}, \&
  {Vlah}}]{HanSelBeu2017}
{Hand}, N., {Seljak}, U., {Beutler}, F., \& {Vlah}, Z. 2017, \jcap, 10, 009,
  \dodoi{10.1088/1475-7516/2017/10/009}

\bibitem[{{Hand} {et~al.}(2012){Hand}, {Addison}, {Aubourg}, {Battaglia},
  {Battistelli}, {Bizyaev}, {Bond}, {Brewington}, {Brinkmann}, {Brown}, {Das},
  \& {Dawson}}]{HanAddAub2012}
{Hand}, N., {Addison}, G.~E., {Aubourg}, E., {et~al.} 2012, Physical Review
  Letters, 109, 041101, \dodoi{10.1103/PhysRevLett.109.041101}

\bibitem[{{Heitmann} {et~al.}(2019{\natexlab{a}}){Heitmann}, {Finkel}, {Pope},
  {Morozov}, {Frontiere}, {Habib}, {Rangel}, {Uram}, {Korytov}, {Child},
  {Flender}, {Insley}, \& {Rizzi}}]{HeiFinPop2019}
{Heitmann}, K., {Finkel}, H., {Pope}, A., {et~al.} 2019{\natexlab{a}}, \apjs,
  245, 16, \dodoi{10.3847/1538-4365/ab4da1}

\bibitem[{{Heitmann} {et~al.}(2019{\natexlab{b}}){Heitmann}, {Uram}, {Finkel},
  {Frontiere}, {Habib}, {Pope}, {Rangel}, {Hollowed}, {Korytov}, {Larsen},
  {Allen}, {Chard}, \& {Foster}}]{HeiUraFin2019}
{Heitmann}, K., {Uram}, T.~D., {Finkel}, H., {et~al.} 2019{\natexlab{b}},
  \apjs, 244, 17, \dodoi{10.3847/1538-4365/ab3724}

\bibitem[{{Hellwing}(2014)}]{Hellwing2014}
{Hellwing}, W.~A. 2014, ArXiv e-prints.
\newblock \doarXiv{1412.8738}

\bibitem[{{Hellwing} {et~al.}(2017){Hellwing}, {Nusser}, {Feix}, \&
  {Bilicki}}]{HelNusFei2017}
{Hellwing}, W.~A., {Nusser}, A., {Feix}, M., \& {Bilicki}, M. 2017, \mnras,
  467, 2787, \dodoi{10.1093/mnras/stx213}

\bibitem[{{Hoffman} {et~al.}(2016){Hoffman}, {Nusser}, {Courtois}, \&
  {Tully}}]{HofNusCorTul2016}
{Hoffman}, Y., {Nusser}, A., {Courtois}, H.~M., \& {Tully}, R.~B. 2016, ArXiv
  e-prints.
\newblock \doarXiv{1605.02285}

\bibitem[{{Howlett} {et~al.}(2017){Howlett}, {Staveley-Smith}, \&
  {Blake}}]{HowStaBla2017}
{Howlett}, C., {Staveley-Smith}, L., \& {Blake}, C. 2017, \mnras, 464, 2517,
  \dodoi{10.1093/mnras/stw2466}

\bibitem[{{Hudson} {et~al.}(2004){Hudson}, {Smith}, {Lucey}, \&
  {Branchini}}]{HudSmiLuc2004}
{Hudson}, M.~J., {Smith}, R.~J., {Lucey}, J.~R., \& {Branchini}, E. 2004,
  \mnras, 352, 61, \dodoi{10.1111/j.1365-2966.2004.07893.x}

\bibitem[{{Hudson} {et~al.}(1999){Hudson}, {Smith}, {Lucey}, {Schlegel}, \&
  {Davies}}]{HudSmiLuc1999}
{Hudson}, M.~J., {Smith}, R.~J., {Lucey}, J.~R., {Schlegel}, D.~J., \&
  {Davies}, R.~L. 1999, \apjl, 512, L79, \dodoi{10.1086/311883}

\bibitem[{{Jaffe} \& {Kaiser}(1995)}]{JafKai1995}
{Jaffe}, A.~H., \& {Kaiser}, N. 1995, \apj, 455, 26, \dodoi{10.1086/176551}

\bibitem[{{Johnson} {et~al.}(2014){Johnson}, {Blake}, {Koda}, {Ma}, {Colless},
  {Crocce}, {Davis}, {Jones}, {Magoulas}, {Lucey}, {Mould}, {Scrimgeour}, \&
  {Springob}}]{JohBlaKod2014}
{Johnson}, A., {Blake}, C., {Koda}, J., {et~al.} 2014, \mnras, 444, 3926,
  \dodoi{10.1093/mnras/stu1615}

\bibitem[{{Juszkiewicz} {et~al.}(2000){Juszkiewicz}, {Ferreira}, {Feldman},
  {Jaffe}, \& {Davis}}]{JusFerFelJaf2000}
{Juszkiewicz}, R., {Ferreira}, P.~G., {Feldman}, H.~A., {Jaffe}, A.~H., \&
  {Davis}, M. 2000, Science, 287, 109, \dodoi{10.1126/science.287.5450.109}

\bibitem[{{Kaiser}(1987)}]{Kaiser1987}
{Kaiser}, N. 1987, \mnras, 227, 1, \dodoi{10.1093/mnras/227.1.1}

\bibitem[{{Kaiser}(1988)}]{Kaiser1988}
---. 1988, \mnras, 231, 149, \dodoi{10.1093/mnras/231.2.149}

\bibitem[{{Kaiser}(1989)}]{Kaiser1989}
{Kaiser}, N. 1989, in Astrophysics and Space Science Library, Vol. 151, Large
  Scale Structure and Motions in the Universe, ed. M.~{Mezzetti},
  G.~{Giuricin}, F.~{Mardirossian}, \& M.~{Ramella}, 197--212,
  \dodoi{10.1007/978-94-009-0903-8_15}

\bibitem[{{Kashlinsky} {et~al.}(2008){Kashlinsky}, {Atrio-Barandela},
  {Kocevski}, \& {Ebeling}}]{KasAtrKoc2008}
{Kashlinsky}, A., {Atrio-Barandela}, F., {Kocevski}, D., \& {Ebeling}, H. 2008,
  \apjl, 686, L49, \dodoi{10.1086/592947}

\bibitem[{{Kumar} {et~al.}(2015){Kumar}, {Wang}, {Feldman}, \&
  {Watkins}}]{KumWanFelWat2015}
{Kumar}, A., {Wang}, Y., {Feldman}, H.~A., \& {Watkins}, R. 2015, ArXiv
  e-prints.
\newblock \doarXiv{1512.08800}

\bibitem[{{Larson} {et~al.}(2011){Larson}, {Dunkley}, {Hinshaw}, {Komatsu},
  {Nolta}, {Bennett}, {Gold}, {Halpern}, {Hill}, \& {Jarosik}}]{WMAP7}
{Larson}, D., {Dunkley}, J., {Hinshaw}, G., {et~al.} 2011, \apjs, 192, 16,
  \dodoi{10.1088/0067-0049/192/2/16}

\bibitem[{{Linder}(2005)}]{Linder2005}
{Linder}, E.~V. 2005, \prd, 72, 043529, \dodoi{10.1103/PhysRevD.72.043529}

\bibitem[{{Macaulay} {et~al.}(2011){Macaulay}, {Feldman}, {Ferreira}, {Hudson},
  \& {Watkins}}]{MacFelFer2011}
{Macaulay}, E., {Feldman}, H., {Ferreira}, P.~G., {Hudson}, M.~J., \&
  {Watkins}, R. 2011, \mnras, 414, 621,
  \dodoi{10.1111/j.1365-2966.2011.18426.x}

\bibitem[{{Macaulay} {et~al.}(2012){Macaulay}, {Feldman}, {Ferreira}, {Jaffe},
  {Agarwal}, {Hudson}, \& {Watkins}}]{MacFelFer2012}
{Macaulay}, E., {Feldman}, H.~A., {Ferreira}, P.~G., {et~al.} 2012, \mnras,
  425, 1709, \dodoi{10.1111/j.1365-2966.2012.21629.x}

\bibitem[{{Masters} {et~al.}(2006){Masters}, {Springob}, {Haynes}, \&
  {Giovanelli}}]{MasSprHay2006}
{Masters}, K.~L., {Springob}, C.~M., {Haynes}, M.~P., \& {Giovanelli}, R. 2006,
  \apj, 653, 861, \dodoi{10.1086/508924}

\bibitem[{Melott {et~al.}(1998)Melott, Coles, Feldman, \&
  Wilhite}]{MelColFelWil1998}
Melott, A.~L., Coles, P., Feldman, H.~A., \& Wilhite, B. 1998, The
  Astrophysical Journal, 496, L85, \dodoi{10.1086/311248}

\bibitem[{{Nusser}(2014)}]{Nusser2014}
{Nusser}, A. 2014, \apj, 795, 3, \dodoi{10.1088/0004-637X/795/1/3}

\bibitem[{{Nusser}(2016)}]{Nusser2016}
---. 2016, \mnras, 455, 178, \dodoi{10.1093/mnras/stv2099}

\bibitem[{{Nusser} {et~al.}(2011){Nusser}, {Branchini}, \&
  {Davis}}]{NusBraDav2011}
{Nusser}, A., {Branchini}, E., \& {Davis}, M. 2011, \apj, 735, 77,
  \dodoi{10.1088/0004-637X/735/2/77}

\bibitem[{{Nusser} \& {Davis}(2011)}]{NusDav2011}
{Nusser}, A., \& {Davis}, M. 2011, \apj, 736, 93,
  \dodoi{10.1088/0004-637X/736/2/93}

\bibitem[{{Okumura} {et~al.}(2015){Okumura}, {Hand}, {Seljak}, {Vlah}, \&
  {Desjacques}}]{OkuHanSel2015}
{Okumura}, T., {Hand}, N., {Seljak}, U., {Vlah}, Z., \& {Desjacques}, V. 2015,
  \prd, 92, 103516, \dodoi{10.1103/PhysRevD.92.103516}

\bibitem[{{Okumura} {et~al.}(2014){Okumura}, {Seljak}, {Vlah}, \&
  {Desjacques}}]{OkuSelVla2014}
{Okumura}, T., {Seljak}, U., {Vlah}, Z., \& {Desjacques}, V. 2014, \jcap, 5,
  003, \dodoi{10.1088/1475-7516/2014/05/003}

\bibitem[{{Planck Collaboration} {et~al.}(2014){Planck Collaboration}, {Ade},
  {Aghanim}, {Armitage-Caplan}, {Arnaud}, {Ashdown}, {Atrio-Barandela},
  {Aumont}, {Baccigalupi}, {Banday}, \& et~al.}]{Planckparameters2014}
{Planck Collaboration}, {Ade}, P.~A.~R., {Aghanim}, N., {et~al.} 2014, \aap,
  571, A16, \dodoi{10.1051/0004-6361/201321591}

\bibitem[{{Planck Collaboration} {et~al.}(2016){Planck Collaboration}, {Ade},
  {Aghanim}, {Arnaud}, {Ashdown}, {Aubourg}, {Aumont}, {Baccigalupi}, \&
  {Banday}}]{PlanckXXXVII2015}
---. 2016, \aap, 586, A140, \dodoi{10.1051/0004-6361/201526328}

\bibitem[{{Planck Collaboration} {et~al.}(2020){Planck Collaboration},
  {Aghanim}, {Akrami}, {Ashdown}, {Aumont}, {Baccigalupi}, {Ballardini},
  {Banday}, {Barreiro}, {Bartolo}, \& {Basak}}]{PlanckAgh2019}
{Planck Collaboration}, {Aghanim}, N., {Akrami}, Y., {et~al.} 2020, \aap, 641,
  A6, \dodoi{10.1051/0004-6361/201833910}

\bibitem[{{Reid} \& {White}(2011)}]{ReiWhi2011}
{Reid}, B.~A., \& {White}, M. 2011, \mnras, 417, 1913,
  \dodoi{10.1111/j.1365-2966.2011.19379.x}

\bibitem[{{Scoccimarro}(2004)}]{Scoccimarro2004}
{Scoccimarro}, R. 2004, \prd, 70, 083007, \dodoi{10.1103/PhysRevD.70.083007}

\bibitem[{{Scrimgeour} {et~al.}(2016){Scrimgeour}, {Davis}, {Blake},
  {Staveley-Smith}, {Magoulas}, {Springob}, {Beutler}, {Colless}, {Johnson},
  {Jones}, {Koda}, {Lucey}, {Ma}, {Mould}, \& {Poole}}]{ScrDavBla2015}
{Scrimgeour}, M.~I., {Davis}, T.~M., {Blake}, C., {et~al.} 2016, \mnras, 455,
  386, \dodoi{10.1093/mnras/stv2146}

\bibitem[{{Seiler} \& {Parkinson}(2016)}]{SeiPar2016}
{Seiler}, J., \& {Parkinson}, D. 2016, \mnras, 462, 75,
  \dodoi{10.1093/mnras/stw1634}

\bibitem[{{Seljak} \& {McDonald}(2011)}]{SelMcD2011}
{Seljak}, U., \& {McDonald}, P. 2011, \jcap, 11, 039,
  \dodoi{10.1088/1475-7516/2011/11/039}

\bibitem[{{Senatore} \& {Zaldarriaga}(2014)}]{SenZal2014}
{Senatore}, L., \& {Zaldarriaga}, M. 2014, ArXiv e-prints.
\newblock \doarXiv{1409.1225}

\bibitem[{{Song} {et~al.}(2013){Song}, {Nishimichi}, {Taruya}, \&
  {Kayo}}]{SonNisTar2013}
{Song}, Y.-S., {Nishimichi}, T., {Taruya}, A., \& {Kayo}, I. 2013, \prd, 87,
  123510, \dodoi{10.1103/PhysRevD.87.123510}

\bibitem[{{Springob} {et~al.}(2007){Springob}, {Masters}, {Haynes},
  {Giovanelli}, \& {Marinoni}}]{SprMasHay2007}
{Springob}, C.~M., {Masters}, K.~L., {Haynes}, M.~P., {Giovanelli}, R., \&
  {Marinoni}, C. 2007, \apjs, 172, 599, \dodoi{10.1086/519527}

\bibitem[{{Springob} {et~al.}(2009){Springob}, {Masters}, {Haynes},
  {Giovanelli}, \& {Marinoni}}]{SprMasHay2009}
---. 2009, \apjs, 182, 474, \dodoi{10.1088/0067-0049/182/1/474}

\bibitem[{{Springob} {et~al.}(2014){Springob}, {Magoulas}, {Colless}, {Mould},
  {Erdo{\u g}du}, {Jones}, {Lucey}, {Campbell}, \& {Fluke}}]{SprMagCol2014}
{Springob}, C.~M., {Magoulas}, C., {Colless}, M., {et~al.} 2014, \mnras, 445,
  2677, \dodoi{10.1093/mnras/stu1743}

\bibitem[{{Sunyaev} \& {Zeldovich}(1980)}]{SunZel1980}
{Sunyaev}, R.~A., \& {Zeldovich}, I.~B. 1980, \mnras, 190, 413,
  \dodoi{10.1093/mnras/190.3.413}

\bibitem[{{Taruya} {et~al.}(2013){Taruya}, {Nishimichi}, \&
  {Bernardeau}}]{TarnisBer2013}
{Taruya}, A., {Nishimichi}, T., \& {Bernardeau}, F. 2013, \prd, 87, 083509,
  \dodoi{10.1103/PhysRevD.87.083509}

\bibitem[{{Taruya} {et~al.}(2010){Taruya}, {Nishimichi}, \&
  {Saito}}]{TarNisSai2010}
{Taruya}, A., {Nishimichi}, T., \& {Saito}, S. 2010, \prd, 82, 063522,
  \dodoi{10.1103/PhysRevD.82.063522}

\bibitem[{Thomas {et~al.}(2004)Thomas, Melott, Feldman, \&
  Shandarin}]{ThoMelFelSha2004}
Thomas, B.~C., Melott, A.~L., Feldman, H.~A., \& Shandarin, S.~F. 2004, The
  Astrophysical Journal, 601, 28, \dodoi{10.1086/380434}

\bibitem[{{Tonry} {et~al.}(2001){Tonry}, {Dressler}, {Blakeslee}, {Ajhar},
  {Fletcher}, {Luppino}, {Metzger}, \& {Moore}}]{TonDreBla2001}
{Tonry}, J.~L., {Dressler}, A., {Blakeslee}, J.~P., {et~al.} 2001, \apj, 546,
  681, \dodoi{10.1086/318301}

\bibitem[{{Tonry} {et~al.}(2003){Tonry}, {Schmidt}, {Barris}, {Candia},
  {Challis}, {Clocchiatti}, {Coil}, {Filippenko}, {Garnavich}, {Hogan},
  {Holland}, {Jha}, {Kirshner}, {Krisciunas}, {Leibundgut}, {Li}, \&
  {Matheson}}]{TonSchBar}
{Tonry}, J.~L., {Schmidt}, B.~P., {Barris}, B., {et~al.} 2003, \apj, 594, 1,
  \dodoi{10.1086/376865}

\bibitem[{{Tully} {et~al.}(2016){Tully}, {Courtois}, \& {Sorce}}]{CF3}
{Tully}, R.~B., {Courtois}, H.~M., \& {Sorce}, J.~G. 2016, \aj, 152, 50,
  \dodoi{10.3847/0004-6256/152/2/50}

\bibitem[{{Tully} {et~al.}(2013){Tully}, {Courtois}, {Dolphin}, {Fisher},
  {H{\'e}raudeau}, {Jacobs}, {Karachentsev}, {Makarov}, {Makarova},
  {Mitronova}, {Rizzi}, {Shaya}, {Sorce}, \& {Wu}}]{TulCouDol2013}
{Tully}, R.~B., {Courtois}, H.~M., {Dolphin}, A.~E., {et~al.} 2013, \aj, 146,
  86, \dodoi{10.1088/0004-6256/146/4/86}

\bibitem[{{Turnbull} {et~al.}(2012){Turnbull}, {Hudson}, {Feldman}, {Hicken},
  {Kirshner}, \& {Watkins}}]{Turnbull2012}
{Turnbull}, S.~J., {Hudson}, M.~J., {Feldman}, H.~A., {et~al.} 2012, \mnras,
  420, 447, \dodoi{10.1111/j.1365-2966.2011.20050.x}

\bibitem[{{Uhlemann} \& {Kopp}(2015)}]{UhlKop2015}
{Uhlemann}, C., \& {Kopp}, M. 2015, \prd, 91, 084010,
  \dodoi{10.1103/PhysRevD.91.084010}

\bibitem[{{Vlah} {et~al.}(2016){Vlah}, {Castorina}, \& {White}}]{VlaCasWhi2016}
{Vlah}, Z., {Castorina}, E., \& {White}, M. 2016, \jcap, 12, 007,
  \dodoi{10.1088/1475-7516/2016/12/007}

\bibitem[{{Wang} {et~al.}(2018){Wang}, {Rooney}, {Feldman}, \&
  {Watkins}}]{WanRooFel2018}
{Wang}, Y., {Rooney}, C., {Feldman}, H.~A., \& {Watkins}, R. 2018, \mnras, 480,
  5332, \dodoi{10.1093/mnras/sty2224}

\bibitem[{{Watkins} \& {Feldman}(2007)}]{WatFel2007}
{Watkins}, R., \& {Feldman}, H.~A. 2007, \mnras, 379, 343,
  \dodoi{10.1111/j.1365-2966.2007.11970.x}

\bibitem[{{Watkins} \& {Feldman}(2015)}]{WatFel2015a}
---. 2015, \mnras, 450, 1868, \dodoi{10.1093/mnras/stv651}

\bibitem[{{Watkins} {et~al.}(2009){Watkins}, {Feldman}, \&
  {Hudson}}]{WatFelHud2009}
{Watkins}, R., {Feldman}, H.~A., \& {Hudson}, M.~J. 2009, \mnras, 392, 743,
  \dodoi{10.1111/j.1365-2966.2008.14089.x}

\bibitem[{{Wegner} {et~al.}(2003){Wegner}, {Bernardi}, {Willmer}, {da Costa},
  {Alonso}, {Pellegrini}, {Maia}, {Chaves}, \& {Rit{\'e}}}]{WegBerWil2003}
{Wegner}, G., {Bernardi}, M., {Willmer}, C.~N.~A., {et~al.} 2003, \aj, 126,
  2268, \dodoi{10.1086/378959}

\bibitem[{{Willick}(1999)}]{Willick1999}
{Willick}, J.~A. 1999, \apj, 516, 47, \dodoi{10.1086/307108}

\bibitem[{{Zaroubi} {et~al.}(1997){Zaroubi}, {Zehavi}, {Dekel}, {Hoffman}, \&
  {Kolatt}}]{ZarZehDekHof1997}
{Zaroubi}, S., {Zehavi}, I., {Dekel}, A., {Hoffman}, Y., \& {Kolatt}, T. 1997,
  \apj, 486, 21

\bibitem[{{Zhang} {et~al.}(2013){Zhang}, {Pan}, \& {Zheng}}]{ZhaPanZhe2013}
{Zhang}, P., {Pan}, J., \& {Zheng}, Y. 2013, \prd, 87, 063526,
  \dodoi{10.1103/PhysRevD.87.063526}

\bibitem[{{Zheng} {et~al.}(2013){Zheng}, {Zhang}, {Jing}, {Lin}, \&
  {Pan}}]{ZheZhaJin2013}
{Zheng}, Y., {Zhang}, P., {Jing}, Y., {Lin}, W., \& {Pan}, J. 2013, \prd, 88,
  103510, \dodoi{10.1103/PhysRevD.88.103510}

\end{thebibliography}
\bibliographystyle{aasjournal}

\end{document}